\documentstyle[twocolumn,eqsecnum,prb,aps]{revtex}

\begin{document}
\draft

\title{Effects of gap anisotropy upon the electronic structure
             around a superconducting vortex}
\author{Nobuhiko Hayashi, Masanori Ichioka, and Kazushige Machida}
\address{Department of Physics, Okayama University,
         Okayama 700, Japan}
\date{May 2, 1997}
\maketitle

\begin{abstract}
   An isolated single vortex is considered within the framework of the
quasiclassical theory.
   The local density of states around a vortex is calculated in a
clean type II superconductor with an anisotropy.
   The anisotropy of a superconducting energy gap is
crucial for bound states around a vortex.
   A characteristic structure of the local density of states,
observed in the layered hexagonal
superconductor 2$H$-NbSe$_2$ by scanning tunneling microscopy (STM),
is well reproduced if one assumes an {\it anisotropic} $s$-wave gap
in the hexagonal plane.
   The local density of states (or the bound states) around the vortex
is interpreted in terms of quasiparticle trajectories
to facilitate an understanding of the rich electronic structure observed in
STM experiments.
   It is pointed out that further fine structures and
extra peaks in the local density of states should be observed by STM.
\end{abstract}



\section{Introduction}
\label{sec:intro}

   Recently, the existence of an anisotropy of a superconducting energy gap 
has attracted a great deal of attention in various superconductors such as
heavy Fermion, organic, and high $T_c$ compounds.
   On the other hand, the electronic structure around vortices is
a fundamental problem
on the physics of both conventional
and unconventional superconductors.
   In this paper, we discuss effects of the gap anisotropy upon 
the electronic structure around a vortex, i.e., the bound states
around an isolated vortex in clean type II superconductors.

   Theoretically, the bound states around a vortex was 
discussed in 1964 by Caroli, 
de Gennes, and Matricon,\cite{caroli} who considered
a single vortex in an isotropic $s$-wave superconductor.
   After this work, several theorists studied
the electronic structure
around vortices.\cite{leadon,bardeen,kramer,watts}
   Experimentally, however, until the following success by Hess
{\it et al.},
there had existed for a long time
no experiments which could
directly study the electronic structure around vortices.\cite{khurana}

   In 1989, a novel experimental method, scanning tunneling microscopy (STM), 
opened up a way to the study of the electronic structure around 
vortices in type II superconductors.\cite{khurana,hess-text}
   Using the STM method, Hess {\it et al.}\cite{hess89} succeeded in
measuring spatially
resolved excitation spectra, i.e., local density of states (LDOS) around 
a vortex.
   They investigated the bound states around a vortex
in the layered hexagonal compound 2$H$-NbSe$_2$ ($T_c$=7.2~K), and found
a striking zero-bias peak at the vortex center.
   The same peak and its collapse upon substituting  Ta for Nb as impurities
in NbSe$_2$ were also observed by Renner {\it et al.}\cite{renner}
   Several new theoretical studies of the electronic structure around
a vortex
\cite{overhauser,daemen,shore,gygi90b,gygi91,klein89,klein90,ullah,pottinger}
were prompted by the success of the STM experiment
by Hess {\it et al.}\cite{hess89}
   Some of these theories\cite{shore,gygi91,klein89,klein90,ullah} predicted
that the zero-bias peak should split into two, i.e., into
positive and negative bias voltage peaks, if spectra 
are taken at some distance from the vortex center (see, for instance, 
Fig. 3 in Ref.\ \onlinecite{shore}).
   This splitting indicates that quasiparticles
of the vortex bound states
with finite angular momentum
are distributed circularly, 
and circulate farther 
away from the core center as they have higher energy.
The predicted splitting was actually confirmed in 
an experiment.\cite{hess90}

   However, a mystery also emerged.
   In the above experiment, Hess {\it et al.}\cite{hess90} not only 
confirmed the splitting, but also found that the LDOS around the vortex was 
shaped like ``star'' at a fixed energy and its orientation was dependent 
on the energy, i.e., the sixfold star shape rotates as 
the bias voltage varies. (Fig. 4 in Ref.\ \onlinecite{hess90}.)
   Soon after this observation was made, Gygi and Schl\"uter\cite{gygi90l}
proposed an explanation for this 30$^{\circ}$ rotation of the star-shaped LDOS.
   On the basis of a sixfold perturbation, 
they explained that the two states, 
i.e., the lower and higher energy stars were interpreted as
bonding or antibonding states.\cite{gygi90l}
   Although they explained certain aspects of the observation,
the following features of the star-shaped LDOS
observed in later experiments\cite{hess91c,hess93,hess-text,hess91b}
could not be sufficiently understood by this perturbation scheme.
   In the intermediate energy, a ``ray'' of the star splits into a pair of
nearly parallel rays.\cite{hess91c,hess93}
(Fig. 1 in Ref.\ \onlinecite{hess91c} or
Fig. 1 in Ref.\ \onlinecite{hess93}.)
  The zero-bias peak in the spectral evolution along a radial line
from the vortex center does not split into two subpeaks observed in
the earlier experiment,
but into three or more ones.\cite{hess-text,hess91b}
(Figs. 9 and 10 in Ref.\ \onlinecite{hess-text}, or
Fig. 6 in Ref.\ \onlinecite{hess91b}.) 

   Specifically, the characteristic features
of the LDOS observed
in NbSe$_2$\cite{hess90,hess91c,hess93,hess-text,hess91b}
are summarized in detail as follows, when the magnetic field $H$ is applied 
perpendicular to the hexagonal plane:
   (1) The LDOS for quasiparticle excitations has a sixfold
star shape centered at the vortex center.\cite{hess90}
   (2) The orientation of this star depends on the energy.
     At zero bias, a ray of the star extends away from
the $a$ axis in the hexagonal plane of NbSe$_2$.
   Upon increasing the bias voltage, the star rotates
by 30$^{\circ}$. 
   (3) In the intermediate bias voltage, a ray splits into
a pair of nearly parallel rays,
keeping its direction fixed.\cite{hess91c,hess93}
   (4) In the spectral evolution which crosses the vortex center,
there exist inner peaks
in addition to the outer peaks which evolve from the zero bias peak at
the vortex center
into the bulk BCS like gap edges far from the vortex.\cite{hess-text,hess91b}
   The inner peaks vary with the angle of the direction in which
the spectral evolution is taken.
These important and interesting observations (1)--(4) remain unexplained.

   Quite recently, motivated by a possibility of
a $d$-wave superconductivity in high $T_c$ cuprates, 
Schopohl and Maki\cite{schopohl,maki} studied
the electronic structure around a vortex in a $d$-wave superconductor.
   On the basis of the quasiclassical
Green's function theory,\cite{eilen,LO,serene}
the LDOS around a single vortex was calculated in a superconductor with
a $d$-wave energy gap.
   They found that the LDOS exhibits a characteristic fourfold
structure in the $d$-wave gap case, which is contrasted with 
the isotropic $s$-wave gap case
(a circularly  symmetric LDOS).\cite{schopohl,maki}
   A gradual 45$^{\circ}$ rotation of this fourfold LDOS 
   as the energy changes was later
reported by the present authors.\cite{oka}
   We note that this rotation is similar to that observed
in NbSe$_2$.\cite{thanks}

   In the context described above, we have investigated the 
electronic structure around the vortex observed in NbSe$_2$.
   We find that the rich structure of the LDOS observed in the STM 
experiments\cite{hess90,hess91c,hess93,hess-text,hess91b} results
mainly from a superconducting gap anisotropy.
   Assuming an {\it anisotropic} $s$-wave gap analogous to the $d$-wave one,
we are able to obtain results favorably comparable with the experiments.
   In a previous short paper,\cite{haya} we enumerated the following 
items as the possible origin of the rich structure of the LDOS:
(a) the effect of an anisotropic superconducting energy gap, 
(b) the effect of nearest-neighbor vortices, i.e., 
the effect of the vortex lattice,
and (c) the effect of the anisotropic density of states at the Fermi surface.
   It is the purpose of the present paper to discuss the gap effect (a)
in more detail.
   As for the item (b), i.e., the vortex lattice effect,
we gave a detailed report in Ref.\ \onlinecite{oka2}.
   A detailed report for (c) the effect of the anisotropic density of states
at the Fermi surface is given elsewhere.\cite{hayashi}

   To date, NbSe$_2$ has been the only compound in which
the electronic structure around vortices was thoroughly
investigated by STM.
   In this paper, we concentrate our attention on the LDOS observed
in NbSe$_2$ as a typical example of a type II superconductor.
   However, the essence of the present considerations
is equally applicable to other type II superconductors in general.
   The LDOS around a vortex reflects the internal 
electronic structure of the vortex,
and an understanding of this structure is important in
elucidating dynamical properties of vortices as well as static ones.

   We consider the case of an isolated static vortex under a magnetic field
applied parallel to the $c$ axis (or $z$ axis).
   We restrict ourselves to a two dimensional system, i.e.,
assume a two dimensional Fermi surface neglecting a small warping
of the Fermi surface along the $c$ axis,
which is appropriate to layered superconductors such as
NbSe$_2$.\cite{corcoran,tanaka}

   In Sec.\ \ref{sec:eilen}, we describe the quasiclassical theory
we use for the study of the vortex.
   Section\ \ref{sec:gap} is devoted to the calculations of 
the LDOS
around a vortex under the influence of 
the gap anisotropy.
   In Sec.\ \ref{sec:trajectry}, we interpret the resultant LDOS
in terms of quasiparticle trajectories.
   The summary and discussions are given in Sec.\ \ref{sec:summary}.

\section{The Quasiclassical Theory}
\label{sec:eilen}
   To investigate the LDOS around a vortex, we use
the quasiclassical Green's function theory.\cite{eilen,LO,serene}
   The quasiclassical theory is a very powerful method,
especially for spatially inhomogeneous systems such as
surfaces\cite{matsumoto,buchholtz}
and vortices.\cite{ours,rainer}
   Furthermore, one can easily treat
a superconducting gap anisotropy as well as 
the Fermi surface anisotropy in the quasiclassical theory.
   We consider the transportlike 
Eilenberger equation for the
quasiclassical Green's function
%
\begin{equation}
{\hat g}(i\omega_n,{\bf r},{\bar{\bf k}})=
-i\pi
\pmatrix{
g(i\omega_n,{\bf r},{\bar{\bf k}}) & if(i\omega_n,{\bf r},{\bar{\bf k}})\cr
-if^{\dagger}(i\omega_n,{\bf r},{\bar{\bf k}}) &
-g(i\omega_n,{\bf r},{\bar{\bf k}}) \cr
}
\label{eq:2.1}
\end{equation}
in a 2$\times$2 matrix form (for even-parity superconductivity),
namely,
%
\begin{eqnarray}
&i&{\bf v}_{\rm F}({\bar{\bf k}})\cdot
{\bf \nabla}{\hat g}(i\omega_n,{\bf r},{\bar{\bf k}})\nonumber\\
& & { }+ \Bigl[\pmatrix{i\omega_n & -\Delta({\bf r},{\bar{\bf k}})\cr
\Delta^*({\bf r},{\bar{\bf k}})
&-i\omega_n\cr},
\  \ {\hat g}(i\omega_n,{\bf r},{\bar{\bf k}}) \Bigr]
=0.
\label{eq:2.2}
\end{eqnarray}
The Eilenberger equation\ (\ref{eq:2.2})
is supplemented by the normalization condition
%
\begin{equation}
{\hat g}(i\omega_n,{\bf r},{\bar{\bf k}})^2
=-\pi^2{\hat 1}.
\label{eq:2.3}
\end{equation}
   Here $\omega_n = (2n+1)\pi T$ is the Matsubara frequency.
The vector ${\bf r}=(x,y)$ is the center of mass coordinate,
and the unit vector ${\bar{\bf k}}$ represents the relative coordinate
of the Cooper pair.
   The overbar denotes unit vectors. 
   The commutator $[{\hat A},{\hat B}]={\hat A}{\hat B}-{\hat B}{\hat A}$.
   We assume the Fermi velocity ${\bf v}_{\rm F}({\bar{\bf k}})$ is
a function of ${\bar{\bf k}}$
with reflecting the anisotropy of the Fermi surface.
   Since we consider an isolated single vortex in an extreme type II
superconductor where the Ginzburg-Landau parameter $\kappa \gg 1$,
the vector potential
can be neglected in Eq. (\ref{eq:2.2}).

   The Eilenberger equation in the matrix form\ (\ref{eq:2.2})
can be written down to the following equations,
%
\begin{mathletters}
\label{eq:2.4}
\begin{equation}
\Bigl( \omega_n +{v_{\rm F}(\theta) \over 2}{d \over d r_{\|}} \Bigr)
f(i\omega_n,{\bf r},\theta)
= \Delta({\bf r},\theta) g(i\omega_n,{\bf r},\theta),
\label{eq:2.4a}
\end{equation}
\begin{equation}
\Bigl( \omega_n -{v_{\rm F}(\theta) \over 2}{d \over d r_{\|}} \Bigr)
f^\dagger(i\omega_n,{\bf r},\theta) 
= \Delta^\ast({\bf r},\theta) g(i\omega_n,{\bf r},\theta),
\label{eq:2.4b}
\end{equation}
\begin{eqnarray}
v_{\rm F}(\theta){d \over d r_{\|}} g(i\omega_n,{\bf r},\theta)
&=& \Delta^\ast({\bf r},\theta) f(i\omega_n,{\bf r},\theta) \nonumber \\
& & { }-\Delta({\bf r},\theta) f^\dagger(i\omega_n,{\bf r},\theta), 
\label{eq:2.4c}
\end{eqnarray}
\end{mathletters}
which are supplemented by
%
\begin{eqnarray}
g(i\omega_n,{\bf r},\theta)
= &[ 1& - f(i\omega_n,{\bf r},\theta)
f^\dagger(i\omega_n,{\bf r},\theta) ]^{1/2},\nonumber\\
&{\rm Re}&\ g(i\omega_n,{\bf r},\theta) > 0.
\label{eq:2.5}
\end{eqnarray}
   Here, ${\bar{\bf k}} = ( \cos\theta, \sin\theta)$,
%
\begin{eqnarray}
{\bf v}_{\rm F}({\bar{\bf k}})
&=& \bigl(|{\bf v}_{\rm F}(\theta)|\cos\Theta(\theta),
 |{\bf v}_{\rm F}(\theta)|\sin\Theta(\theta) \bigr)\nonumber\\
&=& \bigl( v_{\rm F}(\theta)\cos\Theta(\theta),
 v_{\rm F}(\theta)\sin\Theta(\theta) \bigr),
\label{eq:2.6}
\end{eqnarray}
and the following coordinate system is taken:
$\bar{\bf u}=\cos\Theta \bar{\bf x}+\sin\Theta \bar{\bf y}$, 
$\bar{\bf v}=-\sin\Theta \bar{\bf x}+\cos\Theta \bar{\bf y}$, 
and thus a point ${\bf r}=x \bar{\bf x}+y \bar{\bf y}$ 
is denoted as 
${\bf r}=r_{\|} \bar{\bf u}+r_\perp \bar{\bf v}$.
   The center of a vortex line is situated at the origin ${\bf r}=(0,0)$.
   The angle $\theta$, i.e., the direction of
${\bar{\bf k}}$ is measured from
the $a$ axis (or $x$ axis) in the hexagonal plane of NbSe$_2$.
   If one considers a cylindrical Fermi surface
with anisotropic Fermi velocity,
then ${\bf v}_{\rm F}({\bar{\bf k}})
= v_{\rm F}(\theta){\bar{\bf k}}
= \bigl(v_{\rm F}(\theta)\cos\theta, v_{\rm F}(\theta)\sin\theta \bigr)$.

   The self-consistent equation is given by
%
\begin{equation}
\Delta({\bf r},\theta)=N_0 
2 \pi T \sum_{\omega_n>0}
\int_0^{2\pi}{d\theta' \over 2\pi} \rho (\theta')
V(\theta,\theta')
f(i\omega_n,{\bf r},\theta') ,
\label{eq:2.7}
\end{equation}
where $N_0$ is the total density of states over the Fermi surface
in the normal state.
   The $\theta$-dependence of the density of states at the Fermi surface
is represented by
%
\begin{equation}
\rho (\theta) =
{1 \over N_0 |{\bar{\bf k}}\cdot{\bf v}_{\rm F}({\bar{\bf k}})| },
\label{eq:2.8}
\end{equation}
which satisfies $\int(d\theta / 2\pi) \rho (\theta) = 1$.
   We assume that the pairing interaction $V(\theta,\theta')$ is separable,
i.e., $V(\theta,\theta')=vF(\theta)F(\theta')$, where
$v$ is the strength of the pairing interaction
and $F(\theta)$ is a symmetry function, e.g.,
$F(\theta)=\cos2\theta$ for a $d$-wave pairing,
$F(\theta)=1$ for an isotropic $s$-wave pairing, etc.

   The pair potential is written as
%
\begin{equation}
\Delta({\bf r},\theta)=\Delta({\bf r})F(\theta).
\label{eq:2.9}
\end{equation}
   To obtain a self-consistent pair potential,
we solve Eqs.\ (\ref{eq:2.4}) and\ (\ref{eq:2.7}) iteratively.
   This computation
is performed after a method of Ref.\ \onlinecite{oka}.
   In the calculation of the pair potential,
we adopt the so-called explosion method\cite{klein87,thuneberg}
to solve Eq.\ (\ref{eq:2.4}).

   The LDOS is evaluated from
%
\begin{eqnarray}
N(E,{\bf r})&=&N_0\int_0^{2\pi}{d\theta \over 2\pi} \rho (\theta)
{\rm Re}\ g(i\omega_n \rightarrow E+i\eta, {\bf r}, \theta)\nonumber\\
&\equiv&\int_0^{2\pi}{d\theta \over 2\pi} \rho (\theta)
N(E, {\bf r}, \theta),
\label{eq:2.10}
\end{eqnarray}
%
where $\eta$ ($>$0) is a small real constant.
   The value of $\eta$ represents the effect of dilute impurities in a rough
approximation\cite{klein90,pottinger}
or other smearing effects.\cite{dynes}
   To obtain
$g(i\omega_n \rightarrow E+i\eta,{\bf r}, \theta)$, 
we have to solve Eq.\ (\ref{eq:2.4}) for $\eta-iE$
instead of the Matsubara frequency $\omega_n$.
   While we succeeded in this calculation in the vortex lattice case with
the explosion method,
a huge computer-running-time for the numerical calculation
was needed in this method.\cite{oka2}
   In the case of the isolated single vortex, however,
it is convenient to
utilize a method of the Riccati equation
developed by Schopohl.\cite{schopohl,maki,schopohl-u}
   The Riccati equation simplifies the numerical computation.

   The Riccati equations\cite{schopohl} are given as
%
\begin{eqnarray}
&v_{\rm F}&(\theta){d \over d r_{\|}} a(\omega_n,{\bf r},\theta)
-\Delta({\bf r},\theta) \nonumber\\
 &&{ }+\Bigl( 2 \omega_n
 + \Delta^\ast({\bf r},\theta)a(\omega_n,{\bf r},\theta)
\Bigr) a(\omega_n,{\bf r},\theta)=0, 
\label{eq:2.11}
\end{eqnarray}
%
%
\begin{eqnarray}
&v_{\rm F}&(\theta){d \over d r_{\|}} b(\omega_n,{\bf r},\theta)
+\Delta^\ast({\bf r},\theta) \nonumber\\
 &&{ }-\Bigl( 2 \omega_n
 + \Delta({\bf r},\theta)b(\omega_n,{\bf r},\theta)
\Bigr) b(\omega_n,{\bf r},\theta)=0.
\label{eq:2.12}
\end{eqnarray}
   Equations\ (\ref{eq:2.11}) and\ (\ref{eq:2.12}) are obtained
by substituting the following
parametrizations\cite{schopohl} into the Eilenberger equation\ (\ref{eq:2.4}),
%
\begin{equation}
f={2 a \over 1+ab}, \quad
f^\dagger = {2 b \over 1+ab}, \quad
g={1-ab \over 1+ab}.
\label{eq:2.13}
\end{equation}
   We  solve
Eqs.\ (\ref{eq:2.11})
and\ (\ref{eq:2.12}) independently 
along the $r_{\|}$-trajectory
where $r_\perp$ is held constant.
   In the isolated single vortex under consideration,
one can integrate Eqs.\ (\ref{eq:2.11}) and\ (\ref{eq:2.12})
using solutions far from the vortex,
%
\begin{eqnarray}
& & { } a_{-\infty}
= {\sqrt{\omega_n^2 +|\Delta(-\infty,r_\perp,\theta)|^2 } 
-\omega_n \over \Delta^\ast(-\infty,r_\perp,\theta)}, \nonumber\\
b_{+\infty}
&=& {\sqrt{\omega_n^2 +|\Delta(+\infty,r_\perp,\theta)|^2 } 
-\omega_n \over \Delta(+\infty,r_\perp,\theta)}
\quad (\omega_n > 0) 
\label{eq:2.14}
\end{eqnarray}
as the initial values, respectively.\cite{schopohl-u}
   There, to remove the exploding solution,
the integral about $a$ is performed from $r_{\|}=-\infty$,
and about $b$ from $r_{\|}=+\infty$.\cite{schopohl-u}
   We numerically integrate the first-order differential
equations\ (\ref{eq:2.11}) and\ (\ref{eq:2.12})
by the adaptive stepsize control Runge-Kutta method.\cite{numerical} 
   The Green's function $g(i\omega_n \rightarrow E+i\eta,{\bf r}, \theta)$
is obtained from Eq.\ (\ref{eq:2.13})
if one solves Eqs.\ (\ref{eq:2.11}) and\ (\ref{eq:2.12})
for $\eta-iE$ instead of $\omega_n$.
   When we solve Eqs.\ (\ref{eq:2.11})
and\ (\ref{eq:2.12}) for $\eta-iE$,
we use the self-consistently obtained pair potential $\Delta({\bf r})$
which is calculated beforehand.

   From now on, the density of states, energies, and lengths
are measured in units of $N_0$,
the uniform gap $\Delta_0$ at the temperature $T=0$, and the coherence
length $\xi_0 = v_{{\rm F}0}/\Delta_0$ ($v_{{\rm F}0}\equiv 1/N_0$),
respectively.

\section{pair potential and local density of states}
\label{sec:gap}
   Before going into technical details, 
we briefly explain our model and its 
parameter involved in connection with  NbSe$_2$,
which is a typical type II $s$-wave superconductor.
   We assume the following model of an anisotropic $s$-wave pairing
in Eq.\ (\ref{eq:2.9}),
%
\begin{equation}
F(\theta)=1 + c_{\rm A} \cos6\theta .
\label{eq:3.1}
\end{equation}
   Here we again stress that the angle $\theta$, i.e., the direction of
${\bar{\bf k}}$ is measured from
the $a$ axis (or $x$ axis) in the hexagonal plane of NbSe$_2$.
   Thus the parameter $c_{\rm A}$ denotes the degree of anisotropy
in the superconducting energy gap.\cite{markowitz,clem0,clem}
   The case $c_{\rm A}=0$ corresponds to a conventional isotropic gap.

   The anisotropic $s$-wave gap is certainly suggested in NbSe$_2$ from
a scanning tunneling spectroscopy (STS) experiment
at zero field.\cite{hess91b}
   The $I$-$V$ tunneling spectrum, observed at the extreme low temperature
$T=50$ mK, indicates a substantial gap anisotropy
(the gap amplitude with the averaged value 1.1 meV
distributes from 0.7 to 1.4 meV, see Fig. 1 in Ref.\ \onlinecite{hess91b}),
which is consistent with the density of states
in the anisotropic $s$-wave gap case [Fig.\ \ref{fig:dos}].
   It is seen from Fig.\ \ref{fig:dos} that the gap edge distributes from
$E=1-c_{\rm A}$ to $1+c_{\rm A}$ in the case of the anisotropic gap.
   Then, the experimental data of STS\cite{hess91b}
indicate that $c_{\rm A}\sim 1/3$.
   Similarly, a nuclear quadrupole resonance, NQR,
experiment\cite{ishida}
in NbSe$_2$
suggests an anisotropic $s$-wave energy gap.
   The temperature dependence of the spin-lattice
relaxation rate $1/T_1$ is well fitted by an anisotropic energy gap model
following Hebel\cite{hebel}
with the value of a parameter $\delta/{\Delta(0)}\sim 1/3$.\cite{ishida}
   Here the broadening in the gap edge, $\delta/{\Delta(0)}$,
of Ref.\ \onlinecite{ishida} corresponds to $\delta/{\eta_{0}(0)}$
of Ref.\ \onlinecite{hebel}.
   This parameter $\delta/{\Delta(0)}$ corresponds well  to our parameter
of the gap anisotropy, $c_{\rm A}$, because both parameters  
$\delta$ and $c_{\rm A}$ yield the broadening in the gap edge.
   We set $c_{\rm A}=1/3$ as a representative case in the following.

   In this paper, we restrict our attention to the 
gap anisotropy effect only, neglecting other effects,
i.e., the vortex lattice effect and the effect of
the anisotropic density of states at the Fermi surface,
to clearly see how the energy gap anisotropy
influences the LDOS.
   We calculate the LDOS in the isolated vortex case assuming an isotropic
cylindrical Fermi surface
(${\bf v}_{\rm F}({\bar{\bf k}})\parallel{\bar{\bf k}}$,
$v_{\rm F}(\theta)=v_{{\rm F}0}$).

\subsection{Pair potential}
\label{subsec:pair}
   In order to calculate the LDOS, we need the self-consistent pair potential
obtained at the temperature, say,  $T=0.1T_c$
($T_c$ is the superconducting transition temperature).
   The self-consistently obtained real-space variation part
of the pair potential,
$\Delta({\bf r})$, certainly exhibits a weak sixfold structure
both in its phase and amplitude, which results
from the anisotropic pairing, Eq.\ (\ref{eq:3.1}).
   This behavior is similar to that of the $d$-wave case,\cite{oka}
but sixfold symmetric here.
   In Fig.\ \ref{fig:pp}, we show a contour plot of the amplitude of
$\Delta({\bf r})$.
   The amplitude $|\Delta({\bf r})|$ is slightly suppressed in the
$x$ axis direction and its equivalent directions.
   As shown in Fig.\ \ref{fig:pp}, the spatial variation of $\Delta({\bf r})$
has weak anisotropy, but is almost circularly symmetric.
   However, the LDOS shows the characteristic sixfold symmetric structure as
mentioned below.

\subsection{Local Density of States}
\label{subsec:nbse2}
   The LDOS calculated using the self-consistent pair potential
has almost the same structure, except for the length scale for its spread,
as that calculated using a test-potential
$\Delta({\bf r})=\Delta(T) \tanh(r/\xi)\exp(i\phi)$ does,
where $\Delta(T)$ is the uniform gap at the temperature $T$,
$\xi=v_{{\rm F}0}/\Delta(T)$, and the cylindrical coordinate system
${\bf r}=(r,\phi)$ is taken.
   That is, the LDOS does not so affected by the weak sixfold symmetric
spatial structure of the real-space variation part of the pair potential,
$\Delta({\bf r})$.
   We have seen the same situation also in the $d$-wave case.\cite{oka}
   It means that a calculated sixfold or fourfold structure
of the LDOS directly
results from the $k$-space variation part
of the pair potential, $F(\theta)$.

   In Fig.\ \ref{fig:star}, we show the LDOS $N(E,{\bf r})$ for several
energies $E$ in the case $c_{\rm A}=1/3$,
calculated by using the self-consistently obtained pair potential.
   It is seen from Fig.\ \ref{fig:star}(a) that the sixfold star centered
at the vortex center is oriented away from
the $x$ axis by 30$^\circ$ for $E=0$.
   Next it is seen from Fig.\ \ref{fig:star}(b) that at the intermediate
energy each ray splits into two parallel rays, keeping its direction.
   This characteristic feature was precisely observed
in the experiment by Hess.\cite{hess91c,hess93}
   With increasing the energy $E$ further, the sixfold star becomes
a more extended one, and its orientation rotates by 30$^\circ$ as seen
from Fig.\ \ref{fig:star}(c).
   Note that in Fig.\ \ref{fig:star}(c) the head of each ray splits in two.
   It coincides with an experimental result (see the STM image
for 0.48 mV in Fig. 1 of Ref.\ \onlinecite{hess91c}).
   In this way, the anisotropic $s$-wave gap model well reproduces
the experimental features
mentioned in Sec.\ \ref{sec:intro}:
(1) the sixfold star shape, (2) the 30$^\circ$
rotation, and especially (3) the split parallel ray structure at the
intermediate energy.
   We refer to Fig. 1 in Ref.\ \onlinecite{haya} where
the density plots of the LDOS compared with the experimental
data are displayed, which is complimentary to Fig.\ \ref{fig:star}
in the present paper.

   Another way to examine the quasiparticle excitations in
the vortex states is to see how the spectrum evolves along radial lines
from the vortex center.
   We show the spectral evolutions
along the radial lines for 30$^{\circ}$  in Fig. \ref{fig:spectral}(a),
15$^{\circ}$ in\ \ref{fig:spectral}(b),
and 0$^{\circ}$ in\ \ref{fig:spectral}(c)
from the $x$ axis.
   The zero-bias peak splits into several peaks in each spectral evolution.
   Cross sections of each spectral evolution at $r=1$ ($r={\sqrt{x^2 + y^2}}$)
are shown in Fig.\ \ref{fig:cs-spectral} to provide 
the identification of each ridge in Fig.\ \ref{fig:spectral}.

   In the calculation of Figs.\ \ref{fig:star} and\ \ref{fig:spectral},
the smearing factor is chosen as $\eta=0.03$,
which well reproduces the STM experimental data.
   It corresponds to the solid lines of Fig.\ \ref{fig:cs-spectral},
where the peaks are labeled $\alpha$--$\varepsilon$.
   The case with smaller smearing effect ($\eta=0.001$) is represented
by the dashed lines in Fig.\ \ref{fig:cs-spectral},
where the spectrum has the sharp peaks labeled as A--E.
   (The structure of these peaks is discussed in the next section.)
   As shown in Fig.\ \ref{fig:cs-spectral},
by increasing the smearing effect, the spectrum of the dashed line
($\eta=0.001$) is reduced to that of the solid line ($\eta=0.03$),
and reproduces the STM experimental data.
   It seems that the LDOS actually observed in STM experiments
is somewhat smeared due to impurities\cite{klein90,pottinger}
or other smearing effects.\cite{dynes}

   In Fig.\ \ref{fig:spectral}(a) (the 30$^{\circ}$ direction),
there exist one peak at $E=0$ and three pairs of peaks.
   The peak at $E=0$ in Fig.\ \ref{fig:spectral}(a)
[the $\varepsilon$ peak in Fig.\ \ref{fig:cs-spectral}(a)]
corresponds to the ray which extends
in the 30$^{\circ}$ direction in Fig.\ \ref{fig:star}(a).
   This peak is referred to as the inner peak
in Refs.\ \onlinecite{hess-text} and\ \onlinecite{hess91b}.
   This inner peak [the $\varepsilon$ peak]
corresponds to also the split parallel ray
in Fig.\ \ref{fig:star}(b) and
the head of the ray which splits into two in Fig.\ \ref{fig:star}(c).
   The inner $\varepsilon$ peak is, therefore,
sensitive to the angle of the radial line,
and splits in a pair of peaks with the variation of the angle
[see Figs.\ \ref{fig:spectral}(b) and\ \ref{fig:spectral}(c)].
   On the other hand, the most inside pair of peaks
in Fig.\ \ref{fig:spectral}(a)
[the $\delta$ peak in Figs.\ \ref{fig:spectral} and\ \ref{fig:cs-spectral}]    
is not sensitive to the angle.
   This peak is referred to as
the outer peak.\cite{hess-text,hess91b}
   The behavior of the calculated inner and outer peaks well coincide with
the experimental result (the experimental finding (4)
in Sec.\ \ref{sec:intro}).
   The positions of the outer $\delta$ and inner $\varepsilon$ peaks
as a function of $r$
are compared with the experimental data in Fig. 3 of Ref.\ \onlinecite{haya}.

   Outside the inner $\varepsilon$ and outer $\delta$ peaks,
extra peaks appear in each
calculated spectral evolution [the $\alpha$, $\beta$, and $\gamma$ peaks
in Figs.\ \ref{fig:spectral} and\ \ref{fig:cs-spectral}].
   The result of the calculation shows that
the extra peaks are relatively sensitive to the angle of the radial line.
   The existence of the extra peaks is characteristic of the
gap anisotropy effect.
   The peaks $\alpha$ and $\beta$ merge into the upper edge of the
energy gap, $1+c_{\rm A}$, far from the vortex.
   These extra peaks have not been noted in experimental data so far.
   While each peak cannot be clearly identified
in experimental data yet,
it seems that there is at least one new line outside
the outer peak in the data.\cite{hess-priv}
   It is expected for future experiments to definitely identify
the extra peaks.

   The dependence of the LDOS on the angle of the radial
line is important, because it gives a detailed
information on the gap anisotropy.
   To see it, we show in Fig.\ \ref{fig:circle}
a spectral evolution from the angle
0$^{\circ}$ to 30$^{\circ}$ along a circle whose radius $r=1$.
   From this, we can see how each peak moves,
and joins up the others
with the variation of the angle.
   As mentioned above,
the $\varepsilon$ peak (that is, inner peak) is sensitive
to the angle $\phi$ of the radial line, and the $\delta$ peak
(outer peak) is insensitive to $\phi$.
   The $\varepsilon$ peak is located at $E=0$ for $\phi=30^{\circ}$.
   When $\phi$ deviates from $30^{\circ}$,
the peak splits into two which are positive and negative energy peaks.
   With decreasing $\phi$ to 0$^{\circ}$, the energy $E$-position of the
$\varepsilon$ peak increases.
   As for the peaks $\alpha$, $\beta$, and $\gamma$,
with decreasing $\phi$ from 30$^{\circ}$ to 0$^{\circ}$,
the $E$-position decreases for the $\gamma$ peak,
increases for the $\beta$ peak,
and is insensitive for the $\alpha$ peak.
   The peaks $\beta$ and $\gamma$ overlap each other for $\phi=30^{\circ}$,
and the peaks $\alpha$ and $\beta$ overlap each other for $\phi=0^{\circ}$
(see also Fig.\ \ref{fig:cs-spectral}).
   Here, we should mention the behavior of the $\gamma$ peak
at $\phi \sim 0^{\circ}$.
   In Fig.\ \ref{fig:circle}, the $\gamma$ peak seems to
join up the angle-insensitive $\delta$ peak near 0$^{\circ}$,
that is, the $\gamma$ peak is buried in the $\delta$ peak
in Figs.\ \ref{fig:spectral}(c) and\ \ref{fig:cs-spectral}(c)
(the 0$^{\circ}$ direction).
   Such a behavior of the $\gamma$ peak intimately relates to
the value of the anisotropic gap parameter, $c_{\rm A}$.
   The above behavior of $\gamma$ is that of the case $c_{\rm A}=1/3$.
   According as $c_{\rm A}$ increases further,
   the position of the $\gamma$ peak at 0$^{\circ}$
shifts to the higher energy side
[see a spectral evolution
shown in Fig. 3(a) of Ref.\ \onlinecite{haya} (the 0$^{\circ}$ direction),
where $c_{\rm A}$ is set to 1/2 and we can see that
a peak line, which corresponds to the present peak $\gamma$
(not denoted explicitly in that figure),
evolves away from the $\delta$ peak line].

\section{Quasiparticle Trajectories}
\label{sec:trajectry}
   In this section, we interpret the behavior of the quasiparticle
bounded around a vortex in terms of the quasiclassical picture.

\subsection{Direction-dependent Local Density of States}
\label{subsec:dp-ldos}
   In the quasiclassical approximation, the equations are independently
given for each direction of $\bar{\bf k}$.
   The Eilenberger equation (or the Riccati equation) for a direction
$\bar{\bf k}$ is independent of those for the other directions.
   The direction-dependent local density of states $N(E,{\bf r},\theta)$
introduced in  Eq.\ (\ref{eq:2.10}) is obtained from the solution of the
equation for the direction $\bar{\bf k}=(\cos\theta,\sin\theta)$.
   The LDOS $N(E,{\bf r})$ is calculated by integrating
the direction-dependent LDOS $N(E,{\bf r},\theta)$
over $\theta$.
   In an isolated vortex state, the structure of
$N(E,{\bf r},\theta)$ was previously investigated
analytically\cite{kramer,klein89,ullah} and numerically.\cite{klein89}
   According to the results of these investigations,
$N(E,{\bf r},\theta)$ has the following structure for low energies below
$\Delta_0$ in the isolated single vortex.\cite{kramer,klein89,ullah}
   (Here, remind ourselves of the notation:
${\bf r}=x \bar{\bf x}+y \bar{\bf y}=r_{\|} \bar{\bf u}+r_\perp \bar{\bf v}$;
$\bar{\bf u}=\cos\theta \bar{\bf x}+\sin\theta \bar{\bf y}$, 
$\bar{\bf v}=-\sin\theta \bar{\bf x}+\cos\theta \bar{\bf y}$.)
   (i) $N(E,{\bf r},\theta)$ as a function of ${\bf r}=(r_{\|},r_\perp)$
vanishes everywhere except on a straight line along which
$r_\perp={\it const.}=r_\perp(E)$.
   This straight line and $r_\perp(E)$ are referred to as ``quasiparticle
path'' and ``impact parameter,'' respectively.
   (ii) Along the line $r_\perp=r_\perp(E)$, $N(E,{\bf r},\theta)$
has a single maximum at $r_{\|}=0$ and decreases exponentially for
$r_{\|} \rightarrow \pm\infty$.
   (iii) The impact parameter $r_\perp(E)$ is a monotonically increasing
function of $E$.
   One defines $E(r_\perp)$ as the energy level of the state on the
quasiparticle path with the impact parameter $r_\perp$.
   In extreme type II superconductors where $\kappa \gg 1$,
$E(r_\perp)$ is determined by the minimum value of the amplitude
of the pair potential on the
quasiparticle path $r_\perp=r_\perp(E)$.
   For the low energy levels, $E(r_\perp)$ is given
by $E(r_\perp)$=${\rm sgn}(r_\perp)|\Delta(r_{\|}=0,r_\perp)|$
in a good approximation.

   On the basis of the above properties (i)--(iii) of the direction-dependent
LDOS $N(E,{\bf r},\theta)$ studied by
Kramer and Pesch,\cite{kramer} Klein,\cite{klein89}
and Ullah {\it et al.},\cite{ullah}
we interpret our result of the preceding section as follows.

   For simplicity, we concentrate our attention to
Eq.\ (\ref{eq:2.11}) as a representative.
   Dividing
Eq.\ (\ref{eq:2.11}) by $F(\theta)$, we rewrite this equation as
%
\begin{eqnarray}
&&{1\over F(\theta) }{d \over d  r_{\|} } a(\omega_n,{\bf r},\theta)
-\Delta({\bf r}) \nonumber\\
 &&{ }+\Bigl( 2 {\omega_n \over F(\theta)}
 + \Delta^\ast({\bf r}) a(\omega_n,{\bf r},\theta)
\Bigr) a(\omega_n,{\bf r},\theta)=0. 
\label{eq:3.2}
\end{eqnarray}
%

   In the case of the isotropic $s$-wave pairing ($F(\theta)=1$),
$N(E,{\bf r},\theta)$ at a fixed energy
has the identical structure for each direction $\theta$
[the items (i), (ii), and (iii)].
   Then the LDOS $N(E,{\bf r})$, obtained by integrating
$N(E,{\bf r},\theta)$ over $\theta$,
exhibits a ``ring'' shaped structure\cite{schopohl} in the real space.
   The impact parameter is the radius of the ring.

   In the case of an anisotropic pairing,
the situation is changed because of the terms which include $F(\theta)$
in Eq.\ (\ref{eq:3.2}).
   According to Eq.\ (\ref{eq:3.2}), both the length scale
in the $r_{\|}$-direction and the energy scale vary with $\theta$,
but otherwise the form of the equation is same as that of
the isotropic $s$-wave case.
   For the direction $\theta$ where $F(\theta)$ is suppressed,
the length of the spreading of $N(E,{\bf r},\theta)$
along the quasiparticle path [note the items (i) and (ii)]
becomes large.
   For the same $\theta$, the effective energy becomes large and then the
impact parameter becomes far from the vortex center [note the item (iii)].

\subsection{Interpretation on the LDOS around a vortex}
\label{subsec:interpretation}
   We show the partly integrated $N(E,{\bf r},\theta)$ in Fig.\ \ref{fig:part},
where the integration is done from $\theta=-30^\circ$ to 30$^\circ$,
and its schematic figure
in Fig.\ \ref{fig:part-sche}, for the pairing of Eq.\ (\ref{eq:3.1}) where
$c_{\rm A}=1/3$.
   Here, to clarify the structure of the LDOS,
a small smearing parameter ($\eta=0.001$) is adopted.
   The peak lines shown in  Fig.\ \ref{fig:part} are composed of the
quasiparticle paths of each direction $\theta$ described above.
   These peak lines can be interpreted as the flows of quasiparticles
shown in Fig.\ \ref{fig:part-sche}.
   It is noted that the trajectories 1 and 2 appear, because $F(\theta)$ is
finite at $\theta=-30^\circ$ and 30$^\circ$, i.e.,
the impact parameter is finite at these angles.
   If $F(\theta)$ has a node, i.e., $c_{\rm A}=1$, the impact parameter is
infinitely far from the vortex center for the quasiparticle path
of the node direction,\cite{scatt}
and the trajectories 1 and 2 disappear.
   In the bound states, the quasiparticles flow along these trajectories.
   We call it ``quasiparticle trajectory.''
   The whole state at a fixed energy
is composed of such flows of quasiparticles along
the quasiparticle trajectories, while the individual
quasiparticle paths of each
direction $\theta$ [the items (i) and (ii)]
could be considered to be the Andreev reflections.

   We show in Fig.\ \ref{fig:360all} the LDOS $N(E,{\bf r})$ obtained by
integrating the direction-dependent LDOS $N(E,{\bf r},\theta)$
over all $\theta$.
   A schematic figure which corresponds to Fig.\ \ref{fig:360all} is shown
in Fig.\ \ref{fig:sche-360all}.
   The peaks which the radial lines cross are labeled A--E there.
   When the energy $E$ elevates,
the scale of the trajectory in Fig.\ \ref{fig:sche-360all}
increases with keeping its structure fixed.
   Therefore, the trajectory has one-to-one correspondence
to the peak of the spectrum of
Figs.\ \ref{fig:spectral}--\ref{fig:circle}.
   The peaks A--E of Fig.\ \ref{fig:sche-360all} precisely 
correspond to those
of Fig.\ \ref{fig:cs-spectral}.
   These peaks are smeared to appear as $\alpha$--$\varepsilon$ peaks
in  Fig.\ \ref{fig:cs-spectral}
(and thus in  Figs.\ \ref{fig:spectral} and\ \ref{fig:circle}).
   The LDOS actually observed in STM experiments is not that shown in
Fig.\ \ref{fig:360all} itself, but somewhat smeared one
[Figs.\ \ref{fig:star},\ \ref{fig:spectral}, and\ \ref{fig:cs-spectral}]
due to 
impurities\cite{klein90,pottinger} or other smearing effects.\cite{dynes}
   Roughly speaking, the peaks A, B, C, D1, and E correspond to
the peaks $\alpha$, $\beta$, $\gamma$, $\delta$, and $\varepsilon$,
respectively.

   The trajectory of Fig.\ \ref{fig:sche-360all} helps us to facilitate
an understanding of the rich structure of the LDOS.
   The trajectories B and C cross each other at the angle $\phi=30^{\circ}$
from the $x$-axis in Fig.\ \ref{fig:sche-360all}.
   Then, the peaks B and C (i.e., $\beta$ and $\gamma$) overlap each other
in Figs.\ \ref{fig:spectral}(a) and\ \ref{fig:cs-spectral}(a).
   The cross of the trajectories A and B at $\phi=0^{\circ}$
in Fig.\ \ref{fig:sche-360all} corresponds to the overlap of the peaks
A and B (i.e., $\alpha$ and $\beta$) in 
Figs.\ \ref{fig:spectral}(c) and\ \ref{fig:cs-spectral}(c).
   When $\phi$ varies from 30$^{\circ}$ to 0$^{\circ}$,
the trajectories C and D1 cross each other in Fig.\ \ref{fig:sche-360all},
where $c_{\rm A}=1/3$.
   It corresponds to the result that the peaks $\gamma$ and $\delta$
interchange their positions between
Figs.\ \ref{fig:cs-spectral}(b) and\ \ref{fig:cs-spectral}(c).
   However, this behavior of $\gamma$ and $\delta$ depends on
the anisotropic gap parameter $c_{\rm A}$ as mentioned at the end of 
Sec.\ \ref{sec:gap}.
   In the case of large $c_{\rm A}$, the trajectories C and D1 does not cross
for $0^{\circ} \leq \phi \leq 30^{\circ}$ in
Fig.\ \ref{fig:sche-360all}.
Even at $\phi=0^{\circ}$, the trajectory D1 is located farther from
the vortex center than the trajectory C, for large $c_{\rm A}$.
   Then, the peak C (i.e., $\gamma$) is located at higher energy
than the peak D1 (i.e., $\delta$) in the spectrum of
Figs.\ \ref{fig:cs-spectral}(c) and\ \ref{fig:circle},
for large $c_{\rm A}$.
   As seen in Fig.\ \ref{fig:cs-spectral}, the peak D2 tends to be buried
in the other peaks, due to the smearing effects.
   However, if the experiment is performed for the weak smearing case,
the peak D2 should be observed as a small peak, which splits from
the peak D1 (i.e., $\delta$) at $\phi=30^{\circ}$
and approaches the peak E (i.e., $\varepsilon$)
with decreasing $\phi$ to 0$^{\circ}$.
   This D2 peak seems to be easily observed for the angle
$0^{\circ} < \phi < 10^{\circ}$.
   We detect a small indication of the D2 peak for this angle region,
if Fig.\ \ref{fig:circle} is enlarged at $\phi \sim 0^{\circ}$.
   The trajectories D1 and E (i.e., $\delta$ and $\varepsilon$)
corresponds to the trajectories 1 and 2 of Fig.\ \ref{fig:part-sche},
which is related to the lower edge of the anisotropic energy gap.
   Therefore, these trajectories disappear for the higher energy,
$(1-c_{\rm A}) < E < (1+c_{\rm A})$.
   The peaks D1 and E (i.e., $\delta$ and $\varepsilon$) merge into
the lower edge of the energy gap at $E=1-c_{\rm A}$ far from the vortex.

\subsection{Flows of quasiparticles around a vortex}
\label{subsec:flow}
   The flows of the quasiparticles mentioned above are quantitatively
represented by the following quantity,
%
\begin{equation}
{\bf I}(E,{\bf r})=\int_0^{2\pi}{d\theta \over 2\pi} \rho (\theta)
{\bf v}_{\rm F}(\theta) N(E, {\bf r}, \theta),
\label{eq:3.3}
\end{equation}
which we tentatively call ``directional local density of states.''
   This directional LDOS corresponds to a quantity obtained by
integrating
``spectral current density'' introduced by Rainer {\it et al.}\cite{rainer}
over $\theta$ (or ${\bf p}_{f}$ in Ref.\ \onlinecite{rainer}).
   The total current density around a vortex is composed of
the spectral current density.\cite{rainer}
   In Figs.\ \ref{fig:arrow}(a),\ \ref{fig:arrow}(b),\ \ref{fig:arrow}(c),
and\ \ref{fig:arrow}(d),
we show the directional LDOS ${\bf I}(E,{\bf r})$
calculated for $E=0.2$, 1.2, 1.4, and 1.6, respectively.
   Here ${\bf I}(E,{\bf r})$ is calculated under the condition considered
in this section, i.e., under the anisotropic gap and
the isotropic cylindrical Fermi surface.
   It is seen from Fig.\ \ref{fig:arrow}(a) that the flow of the quasiparticle
exhibits a sixfold anisotropy resulting from the sixfold LDOS
of the bound states
(Fig.\ \ref{fig:star}(b) and thus Fig.\ \ref{fig:sche-360all}).
   Now, it is of interest to note the flow with an energy near
the upper gap edge,
$E=1+c_{\rm A}$ ($\simeq$ 1.3).
   Comparing Figs.\ \ref{fig:arrow}(b) and\ \ref{fig:arrow}(d),
we can see that
   the quasiparticles
above and below the upper gap edge
flow each other in reverse directions
except in the vicinity of the vortex center.
   It certainly coincides with a result of an analysis based on the
Bogoliubov-de Gennes (BdG) equation.\cite{gygi91}
   This feature should not be influenced by the gap anisotropy.

\section{Summary and Discussions}
\label{sec:summary}
   The LDOS around an isolated single vortex is studied
within the framework of the quasiclassical theory.
   We consider the effect of the anisotropy of the superconducting
energy gap.
   Assuming the anisotropic $s$-wave energy gap in Eq.\ (\ref{eq:3.1}),
we succeed in theoretically reproducing the characteristic structure
of the LDOS observed in STM experiments;
the observed features, i.e.,
the items (1)--(4) for NbSe$_2$ listed in Sec.\ \ref{sec:intro},
are well described in terms of
the anisotropic gap model.
   We point out the existence of the missing peaks
($\alpha$, $\beta$, and $\gamma$)
at the higher energy side 
in the spectral evolution shown
in Figs.\ \ref{fig:spectral}--\ref{fig:circle}, which is expected to be
looked for in a future experiment.
   We also notice the further splitting of the observed broad
peaks as shown, for example, in Fig. 5(b) ($\delta\rightarrow$ D1 and D2).
   These predictions,
which reflect the gap anisotropy,
 may be checked by using a purer sample
at lower temperatures,
because smearing effects, due to lattice defects or thermal broadening,
mask the fine details.
   We attempt to interpret the calculated LDOS in terms of the
quasiparticle trajectory.
   This enables us to thoroughly understand the 
STM results and the internal electronic structure of the vortex.
   In this paper, the value of our parameter
is chosen appropriate for NbSe$_2$.
   However, the essence of the obtained results should be applicable to
other type II superconductors in general
although the degree of the gap anisotropy $c_{\rm A}$
and the symmetry of $F(\theta)$ will be different in each case.
   Even in other anisotropic superconductors,
the explanation in terms of the quasiparticle trajectory would be
helpful to an understanding of the internal electronic structure of vortices.

\subsection{Comparison with other theories and effects of the vortex lattice}
\label{subsec:theories}
   Let us comment on prior works which are
connected with the star-shaped LDOS observed in NbSe$_2$.
   On the basis of a sixfold perturbation, Gygi and Schl\"uter\cite{gygi90l}
explained that the lower and higher energy stars observed by STM
were interpreted as bonding or antibonding states.
   The STM results (1) and (2) listed in Sec.\ \ref{sec:intro}
were able to be explained by this perturbation scheme.
   They adopted a sixfold crystal lattice potential in NbSe$_2$ as
the perturbation.
   Recently, Zhu, Zhang, and Sigrist\cite{zhu} investigated the effect of the
underlying crystal lattice by means of a non-perturbation method,
i.e., a method of diagonalizing a tight-binding
BdG Hamiltonian in a discrete square lattice,
where the crystal lattice potential, i.e.,
the band structure is determined a priori.
   This method supplements the perturbation theory
of Ref.\ \onlinecite{gygi90l}:
the absolute orientation of the star relative to
the underlying crystal lattice was determined.\cite{zhu}

   By this non-perturbation approach,
also a gradual rotation of the star-shaped LDOS
was obtained in the intermediate energy region.\cite{zhu}
   Nevertheless, it is not yet clear whether the crystal lattice effect
is able to reproduce the remaining experimental findings (3) and (4),
i.e., the split parallel ray structure and the behavior of peaks in
the spectral evolutions.
   The model used in Ref.\ \onlinecite{zhu}
is the discrete lattice model, and therefore it is impossible to obtain
detailed spectra, e.g., spectral evolutions along radial lines,
due to the discreteness.
   Hence, it is desired to treat the crystal lattice potential effect
with a non-perturbation method in the continuum limit.

   Now, the crystal lattice potential determines the band structure,
and influences the structure of the Fermi surface.
   The effect of the crystal lattice potential should appear
as the anisotropy of the Fermi surface.
   In our framework, the anisotropy of the Fermi surface
is taken into account by assuming an anisotropic density of states
at the Fermi surface, $\rho(\theta)$,
which appears in the $\theta$-integral of Eq.\ (\ref{eq:2.10}),
and the anisotropic Fermi velocity $v_{\rm F}(\theta)$,
which appears in the Eilenberger (or Riccati) equations.
   The experimental findings (1)--(4)
can be reproduced qualitatively,
if we introduce a large anisotropy in $v_{\rm F}(\theta)$.
   Its details and the comparison of them to the present results
from the gap anisotropy effect are discussed elsewhere.\cite{hayashi}

   Gygi and Schl\"uter considered also the effect of nearest-neighbor
vortices, i.e, that of the vortex lattice.\cite{gygi90l}
   They adopted a sixfold anisotropy of the vector potential
as the vortex lattice effect,
and treated it as the perturbation.
   However, the periodicity of the pair potential is also
important as the effect of the vortex lattice.\cite{klein89,pottinger}
   In extreme type II superconductors such as NbSe$_2$ where $\kappa \gg 1$,
the periodicity of the pair potential is expected to have stronger effects
upon the structure of the LDOS than the anisotropy of the vector potential
does.
   We find in Ref.\ \onlinecite{oka2} that the effect of the periodicity
gives a characteristic sixfold structure to the LDOS.

   This structure of the LDOS which results from
the periodicity of the pair potential appears only at high magnetic fields
such as 1~T for the material parameters appropriate to NbSe$_2$,
where the vortex core regions substantially overlap each other.\cite{oka2}
   At a lower magnetic field such as 0.1~T, the calculated
LDOS reduces to the almost
circular structure.
   On the other hand, the LDOS observed in a STM experiment
exhibits the star-shaped
structure in spite of a low field 0.025~T
(see Fig. 12 in Ref.\ \onlinecite{murray}).
   Therefore, in the case of NbSe$_2$ at low magnetic fields,
we need to consider the effects of anisotropy other than the vortex lattice
effect in order to explain the star-shaped LDOS.
   Both the vortex lattice effect and the anisotropic superconducting
gap one
are important for the star-shaped LDOS observed in NbSe$_2$
at high magnetic fields.
   We expect a future STM experiment to be performed
on isotropic superconducting compounds or metals
to clarify the vortex lattice effect and confirm predictions of
Ref.\ \onlinecite{oka2}.

   In STM experiments on NbSe$_2$, one of the directions
of nearest-neighbor
vortices coincides with the $a$ axis
(see the literature by Hess {\it et al.}
\cite{hess-text,hess89,hess90,hess91c,hess93,hess91b,murray,murray-l}
or Renner {\it et al.}\cite{renner}),
except for extreme low fields.\cite{bolle}
   This experimental fact gives evidence of a correlation of the vortex
lattice with the underlying crystal lattice of NbSe$_2$.
   It was recently found that in $d$-wave superconductors, higher-order
(nonlocal correction) terms in the Ginzburg-Landau equation,
which reflect the fourfold symmetric property of the $d$-wave pairing,
give rise to a preferred direction of
the vortex lattice.\cite{won,eno}
   In NbSe$_2$, the sixfold anisotropic pairing, Eq.\ (\ref{eq:3.1}),
is expected to give rise to the same correlation as the $d$-wave pairing does,
and it may be the origin of the experimental fact mentioned above.
   A possibility of the correlation of the vortex lattice with
the underlying crystal lattice was recently reported also
in a high $T_c$ cuprate.\cite{keimer,renner2}

\subsection{Beyond the quasiclassical approach}
\label{subsec:beyond}
   We mention the LDOS around a vortex in high $T_c$ cuprates.
   It seems from various experiments that
high $T_c$ material is a $d$-wave superconductor.\cite{tsuei}
   A fourfold structure of the LDOS is predicted in $d$-wave superconductors
by theoretical studies based on the quasiclassical
theory.\cite{schopohl,maki,oka}
   The origin of this fourfold structure is same as that discussed in the
present paper for the gap anisotropy.
   Recently, Maggio-Aprile {\it et al.} observed tunneling spectra
around vortices in a high $T_c$ cuprate, YBa$_2$Cu$_3$O$_{7-\delta}$,
with STM.\cite{maggio,levi,renner2}
   However, the spectroscopic images of STM have not
exhibited any sign of a fourfold structure yet.
   We expect further detailed experiments
to observe the fourfold symmetric LDOS structure.

   When we consider the high $T_c$ materials,
the quantum effects should be taken into account.
   The quasiclassical theory is certainly valid only in systems
where the atomic scale spatial variation of the Green's function
can be neglected with respect to
the coherence length scale one.\cite{serene,morita}
   The effects neglected in the quasiclassical theory can be important
in the case of the high $T_c$ cuprate;
the quantization of energy levels of the bound states cannot be treated
by the quasiclassical theory,
and while it is possible in the quasiclassical approximation
to divide the equation
into individual equations for each direction of $\bar{\bf k}$,
it is impossible in the quantum-mechanical limit.
   Although we expect the fourfold structure of the LDOS should be observed
in future experiments,
the above effects may change the situation
in the case of the high $T_c$ cuprate.
   It is certainly desired on the theoretical side
that a fully quantum-mechanical approach
clears up this problem in future.

   As for the fully quantum-mechanical approach, it is needed to solve
the BdG equation without quasiclassical
approximations.
   The BdG equation cannot be written in a local form
in the case of an anisotropic pairing,
and therefore it is difficult to treat this equation in the continuum limit.
   One of the possible approaches to this problem is
the method of diagonalizing a BdG Hamiltonian for
a specific lattice model.\cite{zhu,soininen,wang,feder,himeda}
   In the lattice model, however, the atomic scale variation of wave functions
among the lattice points is uncertain.

   In most superconductors ($\xi \gg 1/k_{\rm F}$),
the atomic scale variation of the wave function
is a redundant information
and can usually be neglected
on the basis of the quasiclassical theory.
   On the other hand, in the high $T_c$ cuprate superconductors,
$k_{\rm F} \xi \simeq \varepsilon_{\rm F}/\Delta_{0} \sim 1$
(the Fermi wave-number and energy are $k_{\rm F}$ and $\varepsilon_{\rm F}$,
respectively),\cite{morita}
and therefore the atomic scale variation and the quantization of bound states
in a vortex may be crucial for the electronic structure around the vortex
in the cuprates.

   The high $T_c$ cuprate is certainly the only superconductor possessed of
a possibility of an experimentally
detectable quantization in the vortex bound states.
   According to Ref.\ \onlinecite{caroli},
a substantial energy quantization
(of the order of $\Delta_{0}^2/\varepsilon_{\rm F} \sim 10$ K) is
expected to exist in the high $T_c$ cuprate.
   However, to the present author's knowledge, the system considered in
Ref.\ \onlinecite{caroli} is an isotropic $s$-wave superconductor and
the mechanism of the quantization in the case of anisotropic pairing
is not yet understood.
   In case of gap node due to anisotropic pairing,
it is expected that the separation of the energy levels becomes small.
   Further experiments, which, e.g., investigate spatial variation of
this quantized bound states in the high $T_c$ cuprates with STM
and then compare its result with the quasiclassical
prediction\cite{schopohl,maki,oka}
in order to clarify how the quantum effects mentioned above
modify the vortex bound states,
are the need for alternative theoretical studies
of the vortex bound states.

\subsection{Concluding remarks}
\label{subsec:remarks}
   The electronic structure of vortices in a compound,
LuNi$_2$B$_2$C, was quite recently investigated
by STM.\cite{dewilde,dewilde2}
   Although no conductance peaks related to localized quasiparticle states
in the vortex core are observed in the experiment, due to a short mean free
path (of the order of the coherence length) and thermal broadening effects
at 4.2~K ($T_c \approx 16~K)$,\cite{dewilde}
a rich (maybe fourfold) structure of the LDOS
such as that discussed in the present paper is expected to be detected
in STM spectra by
lowering the temperature and decreasing impurities or defects.
   If an anisotropic bound states around a single vortex is observed,
it should suggest an anisotropy of the pairing in this compound.
   The direction (in the $k$-space)
in which the superconducting gap is suppressed
corresponds to that (in the real space) of a ray of the LDOS at zero bias. 

   Finally, low-temperature STM is the unique experimental method
which has the ability not only to image the distribution of
the vortex lattice, but also to probe the electronic
structure of individual vortices.
   We expect future STM experiments to be performed in vortex states
on various superconductors
such as organic conductors, high $T_c$ cuprates, heavy Fermion
superconductors (e.g., UPt$_3$), and a recently discovered non-copper-layered
perovskite superconductor, Sr$_2$RuO$_4$\cite{maeno} which has nearly
cylindrical Fermi surfaces\cite{oguchi,singh} and a possibility that
an odd-parity superconductivity would be realized in it.\cite{sigrist,machida2}
   The information on the vortex bound states available from STM spectra
can be one of clues to the pairing. 
The low-temperature STM experiments deserve a great deal of attention.

\acknowledgments

   It is our pleasure to thank H.~F.~Hess for fruitful suggestion
on the present issue,
useful communication of results, and
helpful discussions.
   We are also grateful to N.~Schopohl for sending useful information
on the Riccati equation
and to Ch.~Renner for useful discussion of their experiments.
   We are indebted to the Supercomputer Center of the
Institute for Solid State Physics, University of Tokyo,
for part of the numerical calculations.
   One of us (M.I.) is supported by the Japan Society
for the Promotion of Science for Young Scientists.


\newpage
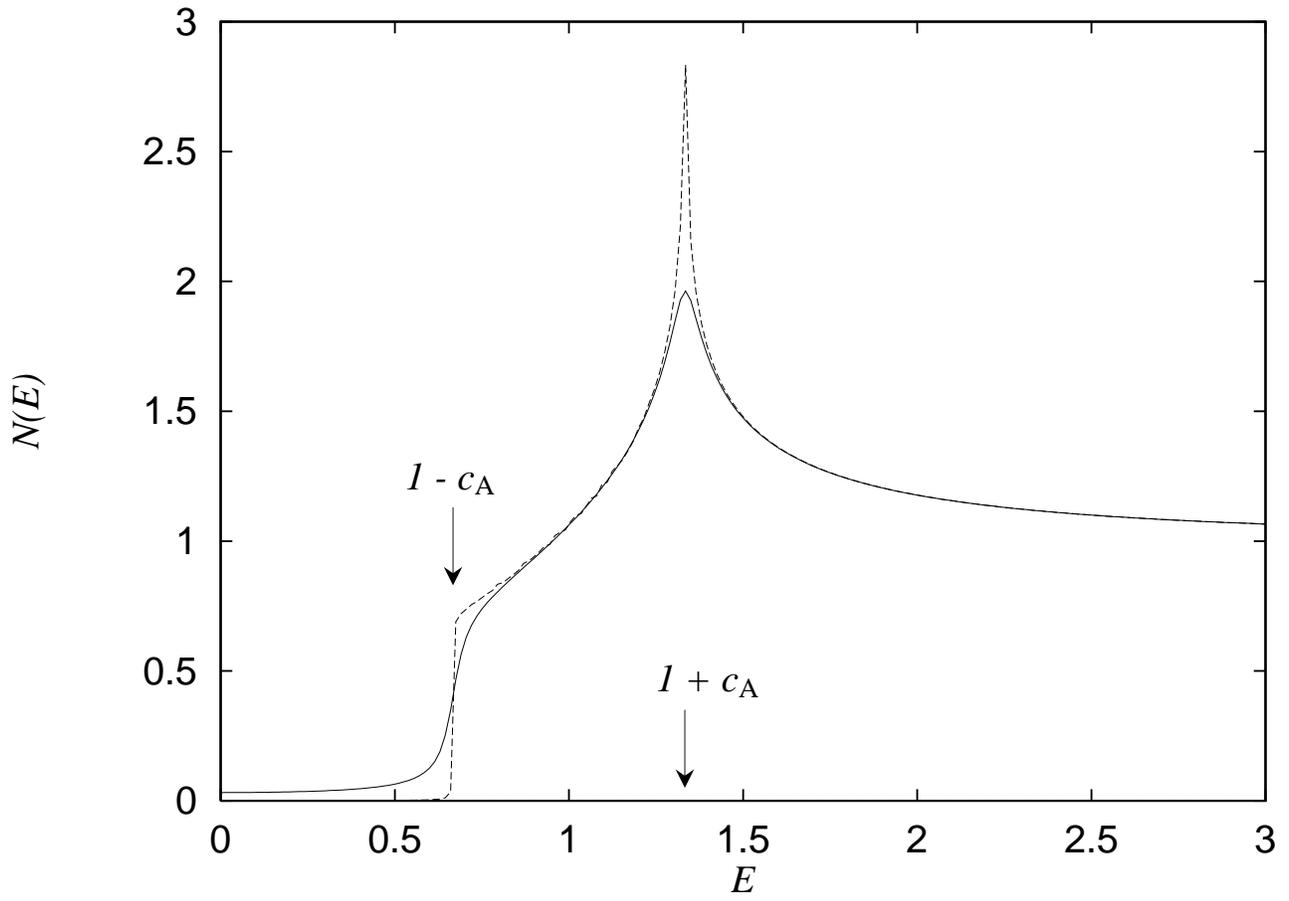
\begin{figure}
\caption{
   The density of states $N(E)$ at zero field,
where smearing parameter
$\eta=0.03$ (solid line) or $\eta=0.001$ (dashed line).
It is calculated for the isotropic cylindrical Fermi surface, and
the degree of the gap anisotropy is $c_{\rm A}=1/3$.
}
\label{fig:dos}
\end{figure}
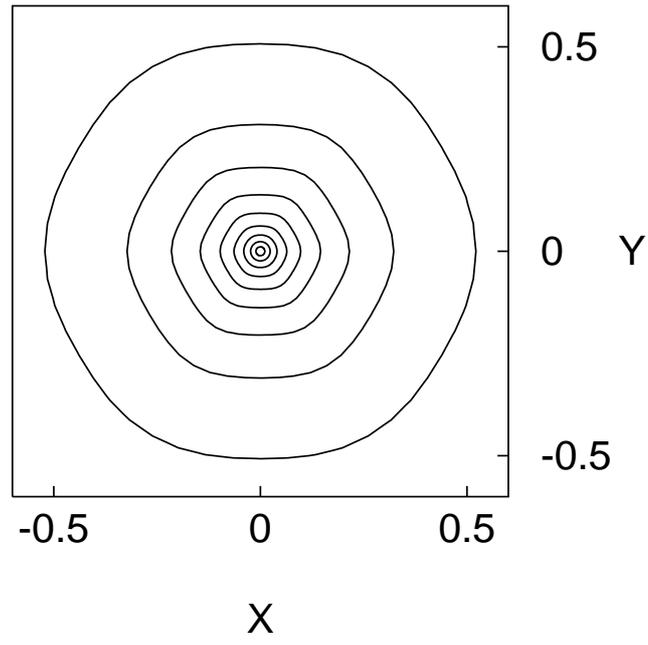
\begin{figure}
\caption{
Contour plot of the amplitude of the real-space variation part of
the pair potential, $|\Delta({\bf r})|$.
From the center, 0.1, 0.2, \ldots, 0.9.
The temperature is $T=0.1T_c$, and
the degree of the gap anisotropy is $c_{\rm A}=1/3$.
}
\label{fig:pp}
\end{figure}
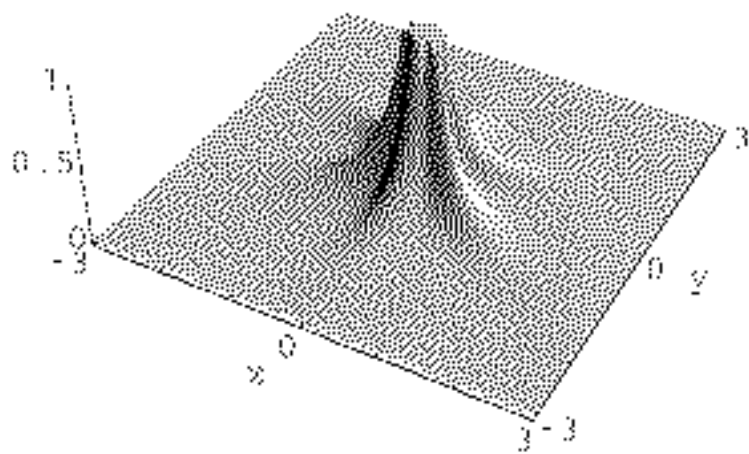
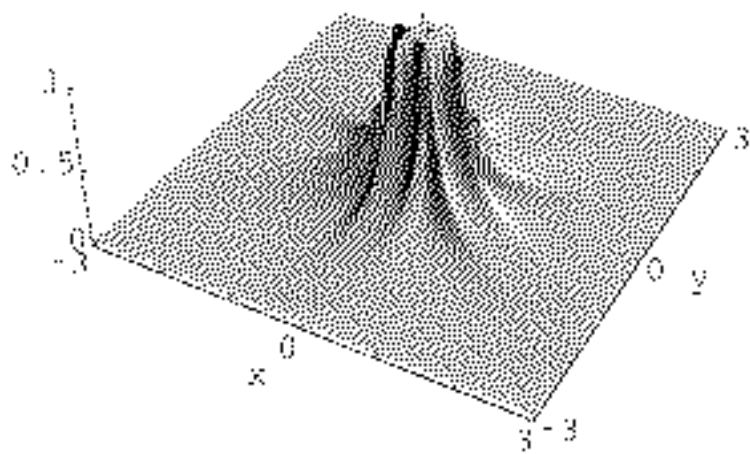
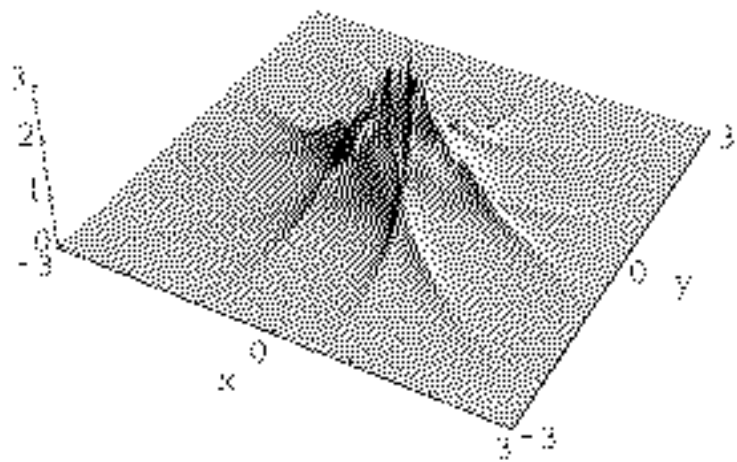
\begin{figure}
\caption{
The LDOS $N(E,{\bf r})$ ($\eta=0.03$)
calculated for the energies $E=0$ (a), 0.2 (b), and 0.4 (c).
Large peaks in the vicinity of the vortex center are truncated
in the figures (a) and (b).
}
\label{fig:star}
\end{figure}
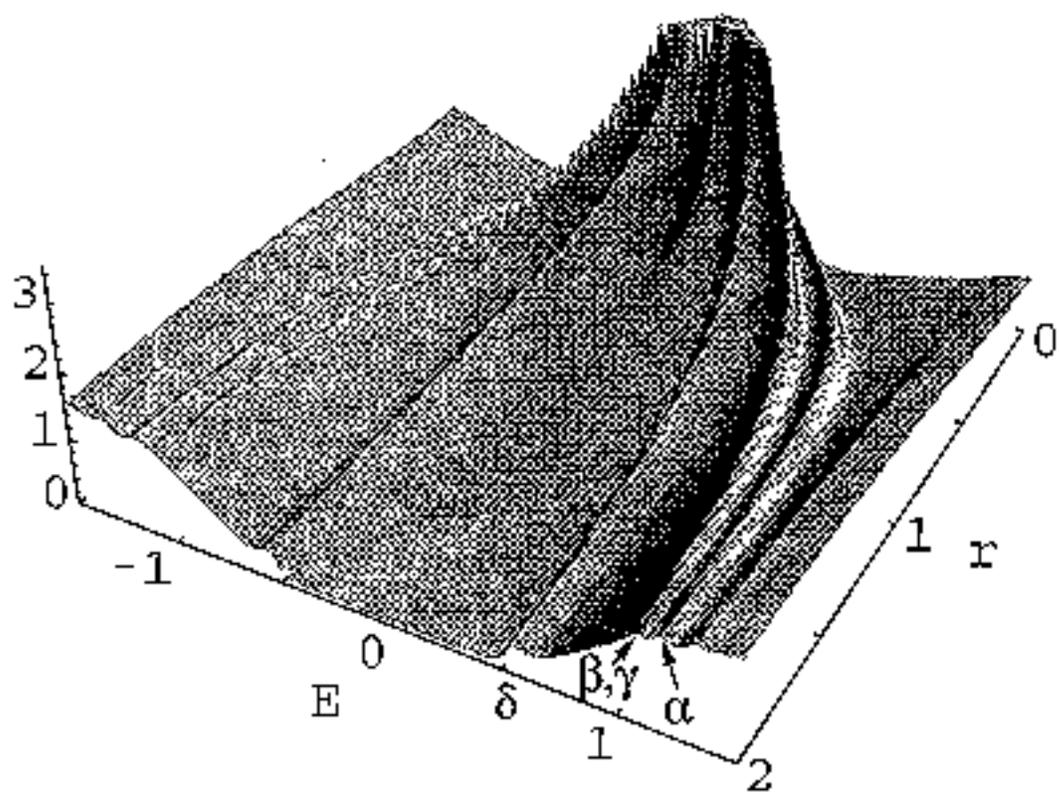
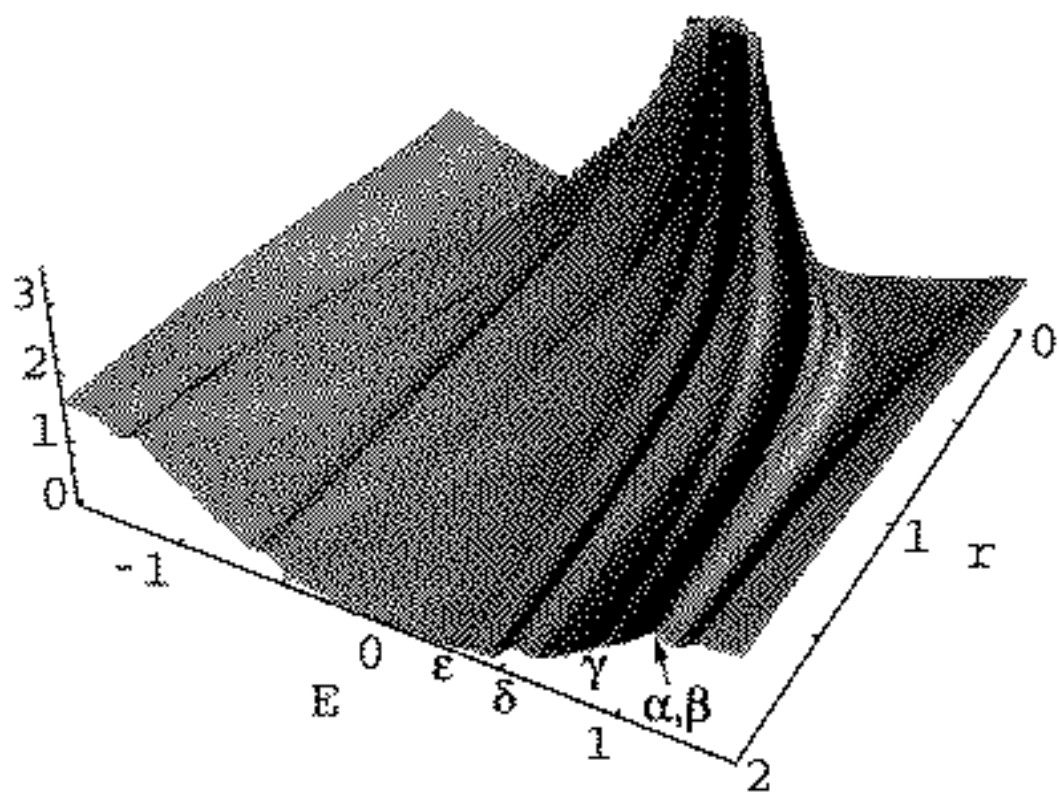
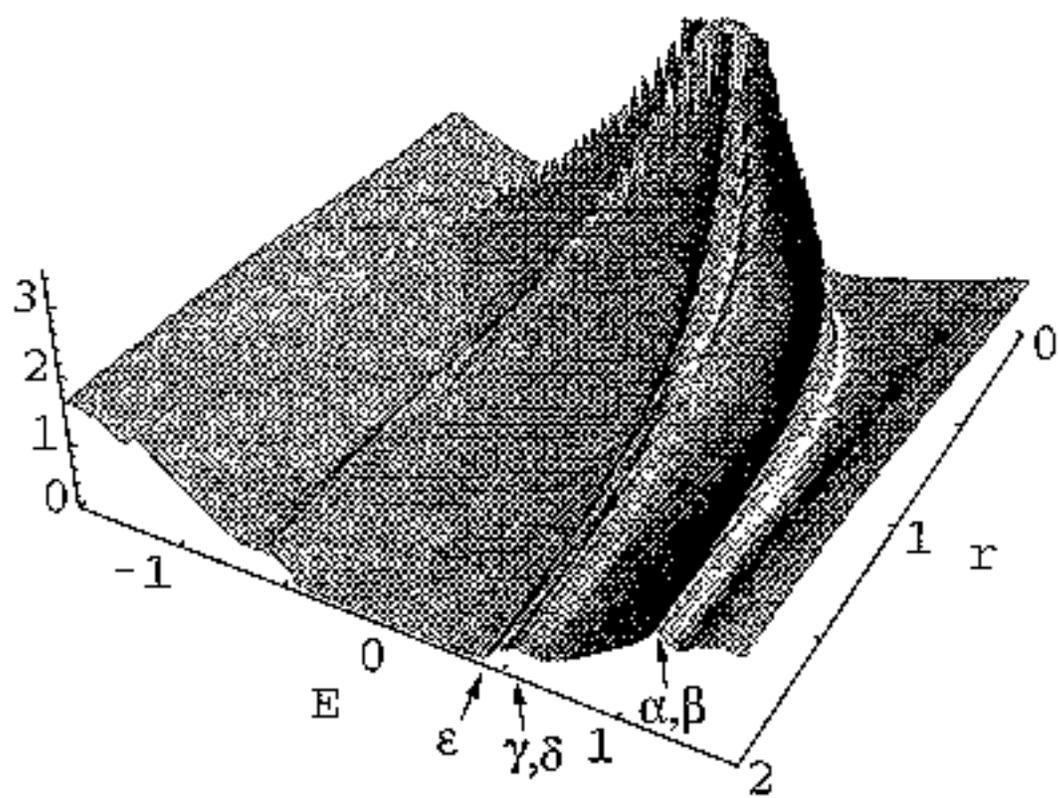
\begin{figure}
\caption{
Spectral evolutions $N(E,{\bf r})$ ($\eta=0.03$)
along radial lines for 30$^\circ$ (a),
15$^\circ$ (b), and 0$^\circ$ (c) from the $x$ axis.
The zero-bias peak is truncated in the figures.
The peak lines in the spectra are labeled $\alpha$--$\varepsilon$.
}
\label{fig:spectral}
\end{figure}
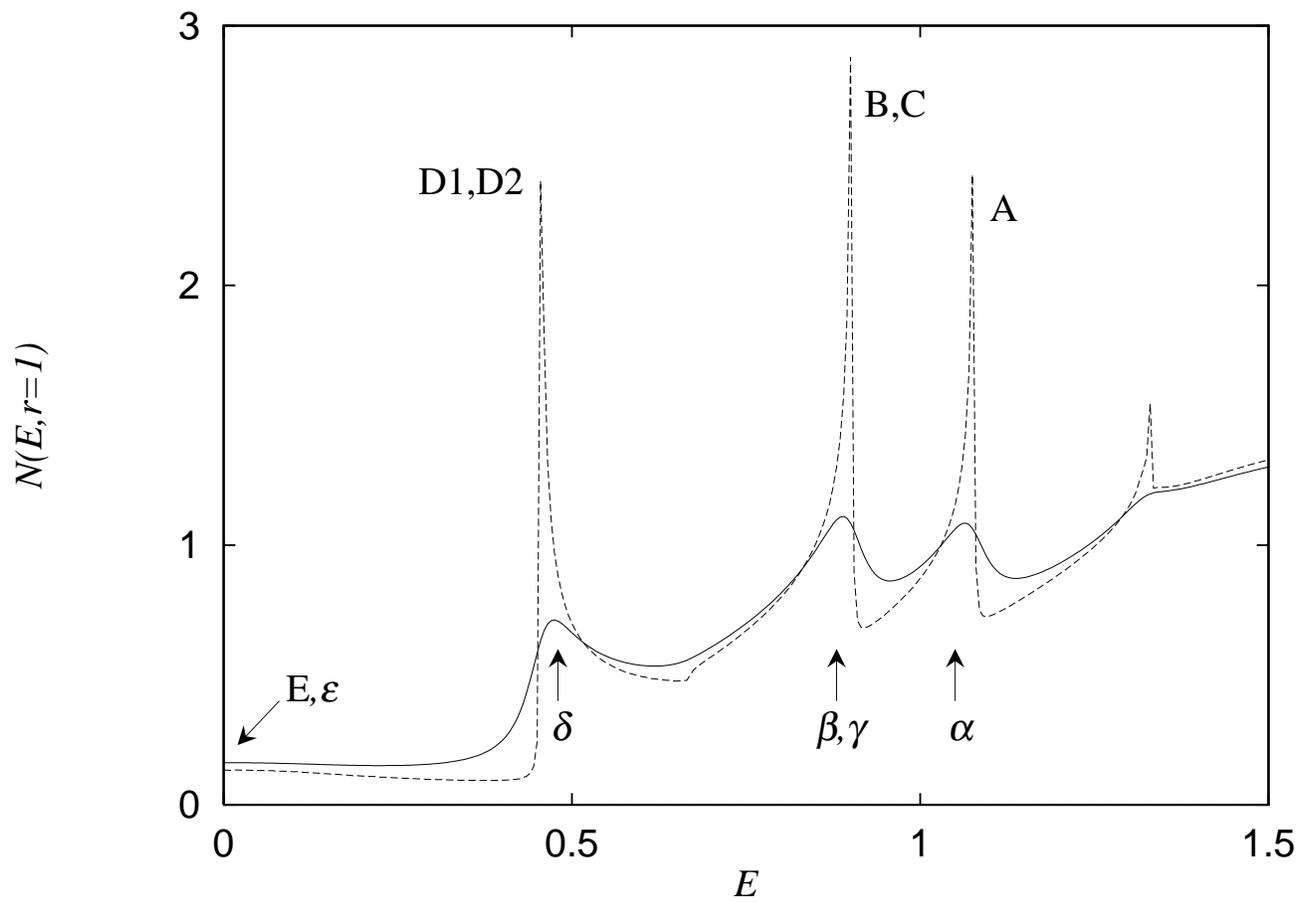
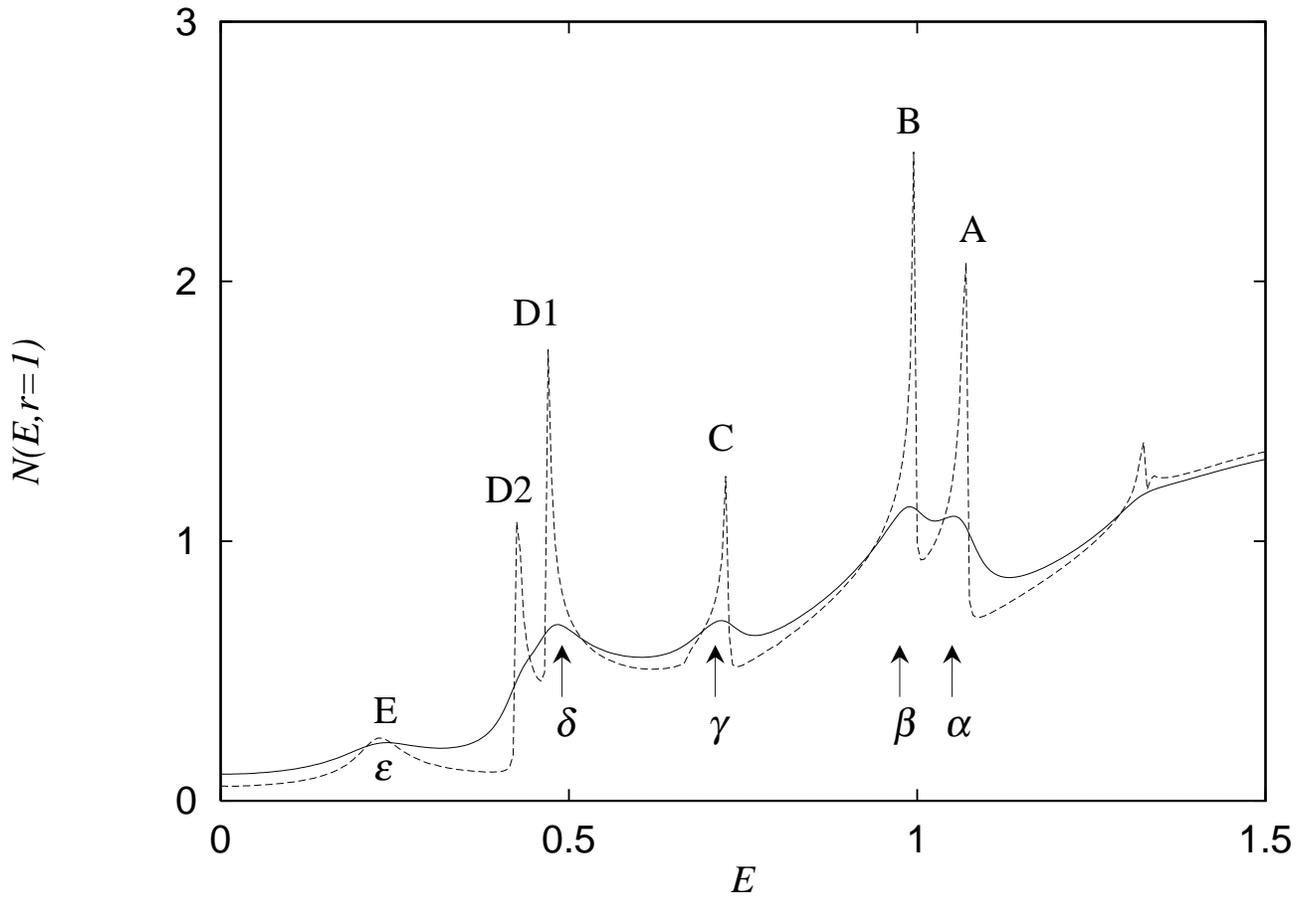
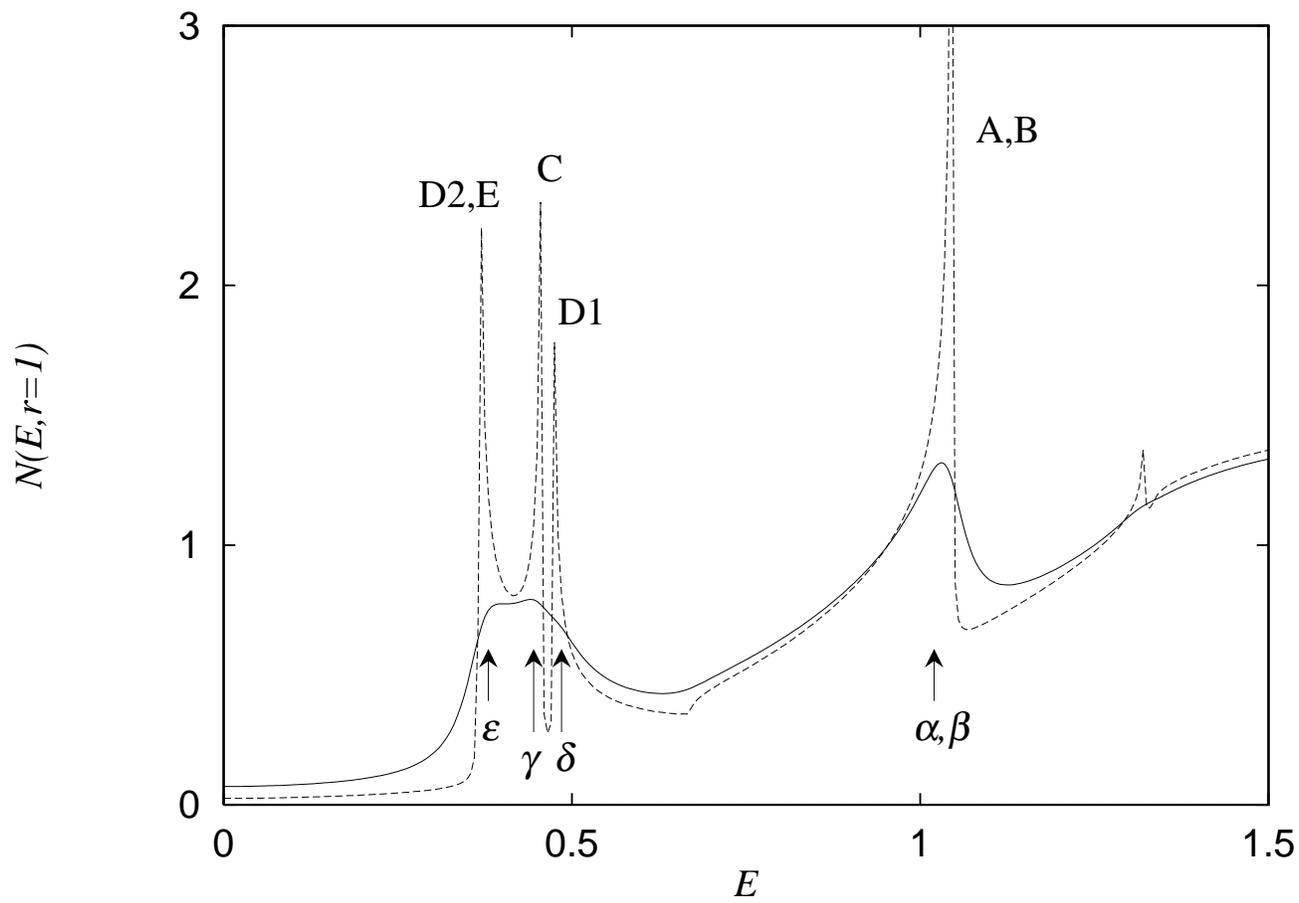
\begin{figure}
\caption{
Cross sections of the spectral evolutions (Fig. 4) 
at the distance from the vortex center, $r=1$.
The directions of each radial line are
30$^\circ$ (a), 15$^\circ$ (b), and 0$^\circ$ (c) from the $x$ axis.
The peaks in the spectra are labeled A--E for the dashed line spectra
($\eta=0.001$) and $\alpha$--$\varepsilon$ for the solid line spectra
($\eta=0.03$).
The labels $\alpha$--$\varepsilon$ correspond to those of Fig. 4.
}
\label{fig:cs-spectral}
\end{figure}
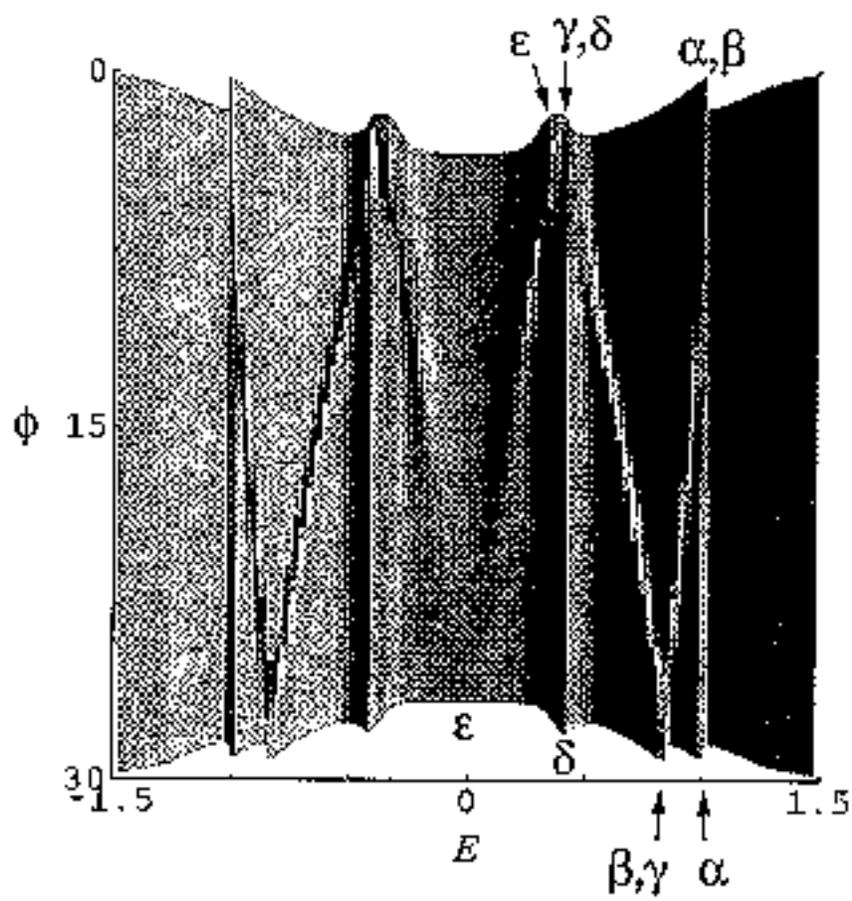
\begin{figure}
\caption{
Spectral evolution $N(E,{\bf r})$ ($\eta=0.03$)
from the angle 0$^{\circ}$ to 30$^{\circ}$ along
a circle whose radius $r=1$.
The center of this circle is situated at the vertex center.
The peaks labeled as $\alpha$--$\varepsilon$ correspond to
those of Figs. 4 and 5.
}
\label{fig:circle}
\end{figure}
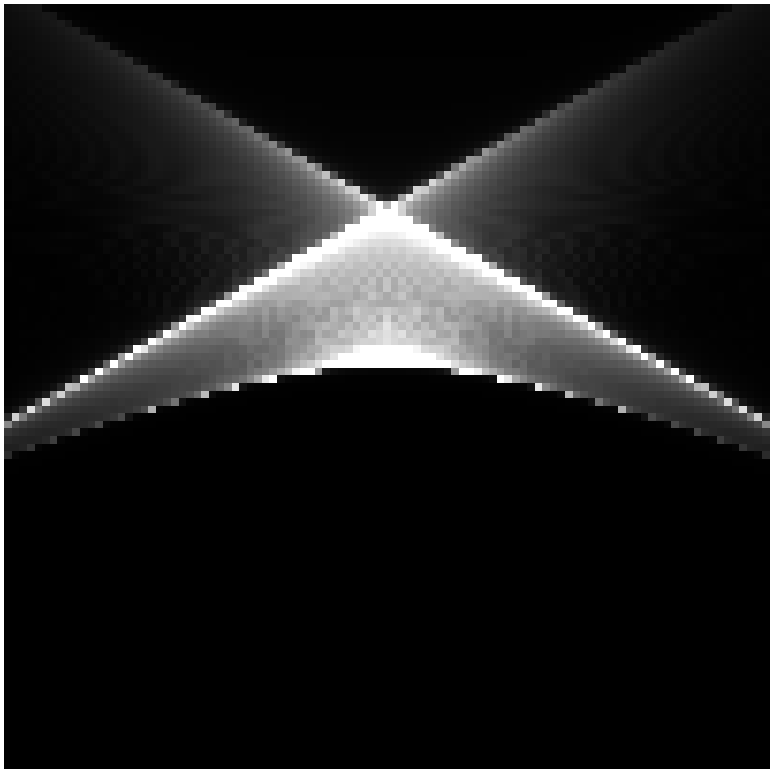
\begin{figure}
\caption{
The direction-dependent LDOS
$N(E,{\bf r},\theta)$ partly integrated from $\theta=-30^\circ$ to 30$^\circ$,
where $E=0.5$, $\eta=0.001$, and
$6\xi_0 \times 6\xi_0$ is shown in the real space.
}
\label{fig:part}
\end{figure}
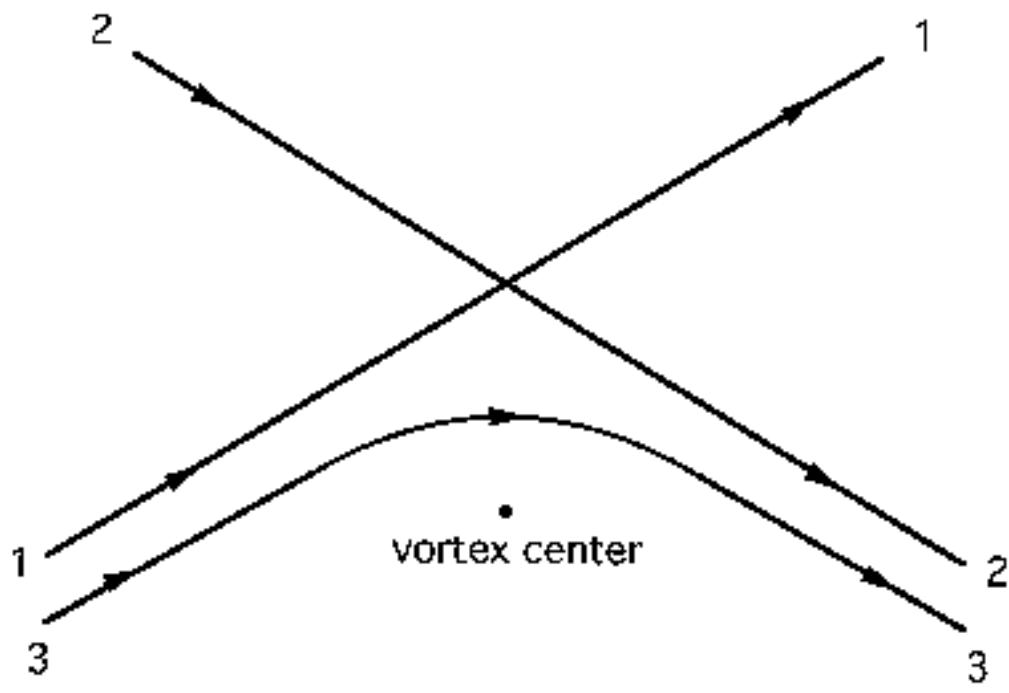
\begin{figure}
\caption{
Schematic flow trajectories of quasiparticles with an energy
$0< E < (1-c_{\rm A})$.
These trajectories correspond to those shown
in Fig. 7.
When $(1-c_{\rm A}) < E < (1+c_{\rm A})$,
the trajectories 1 and 2 disappear and only the trajectory 3 is alive.
}
\label{fig:part-sche}
\end{figure}
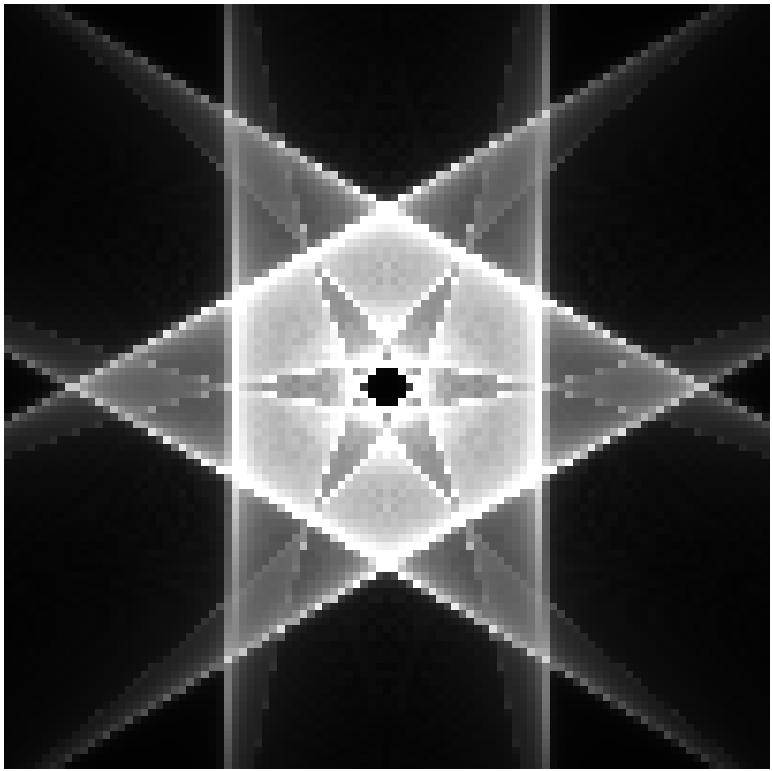
\begin{figure}
\caption{
The LDOS $N(E,{\bf r})$ which is obtained by integrating
the direction-dependent LDOS $N(E,{\bf r},\theta)$ over $\theta$, where
$E=0.5$, $\eta=0.001$, and
$6\xi_0 \times 6\xi_0$ is shown.
}
\label{fig:360all}
\end{figure}
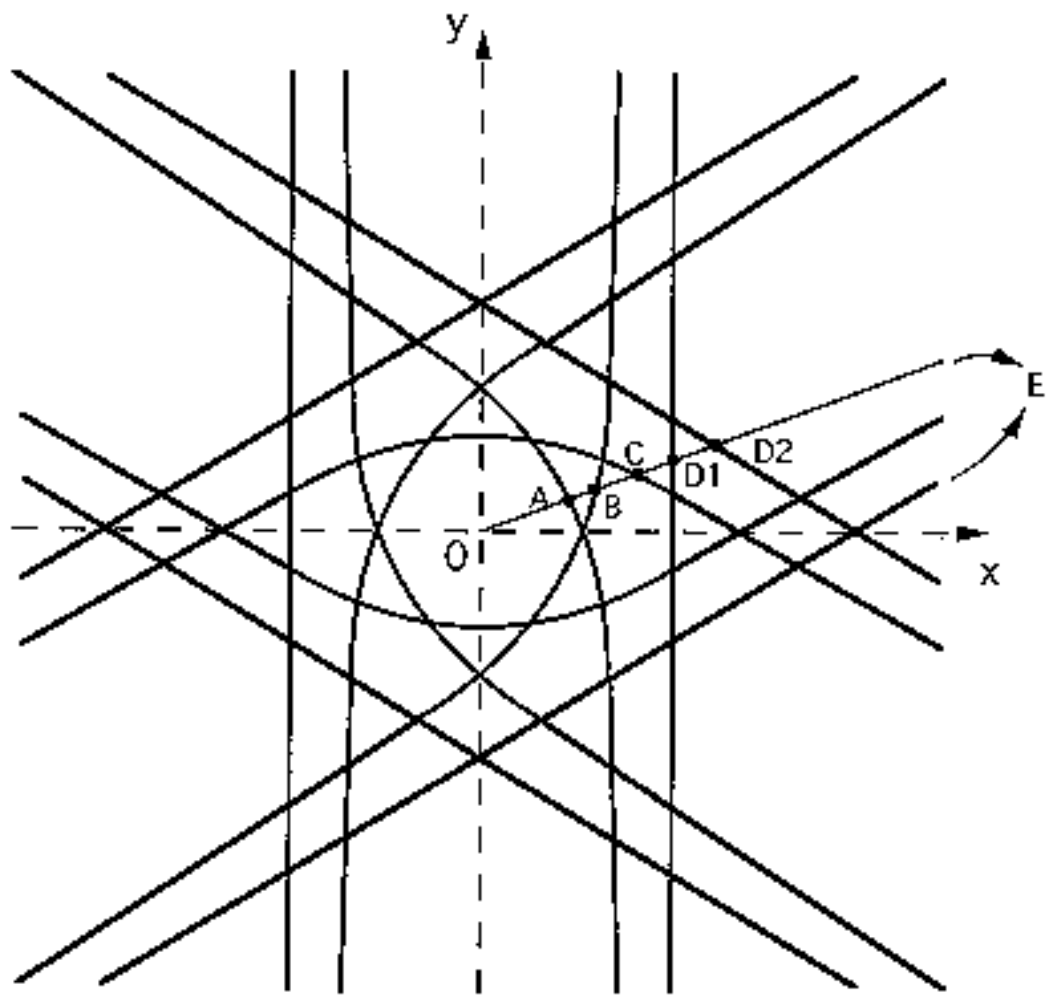
\begin{figure}
\caption{
Schematic figure of the LDOS $N(E,{\bf r})$ for
an energy $0< E < (1-c_{\rm A})$.
Points A--E correspond to the peaks of the dashed line spectra in Fig. 5.
}
\label{fig:sche-360all}
\end{figure}
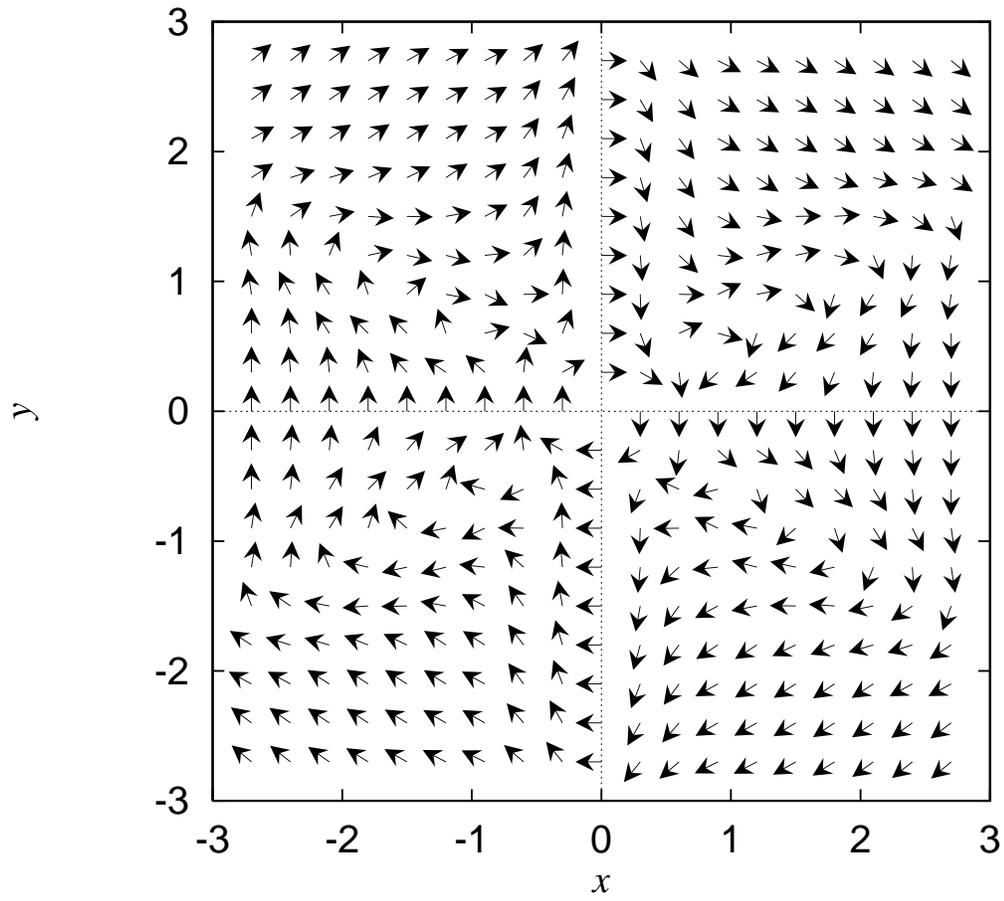
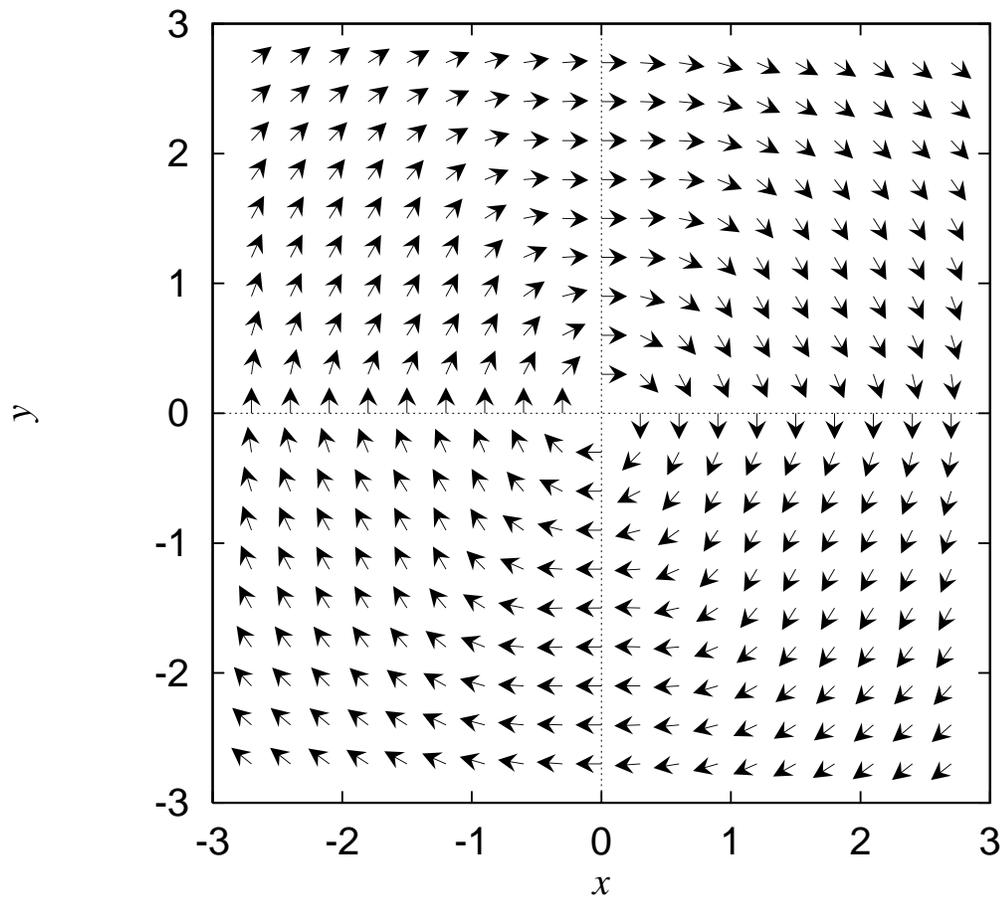
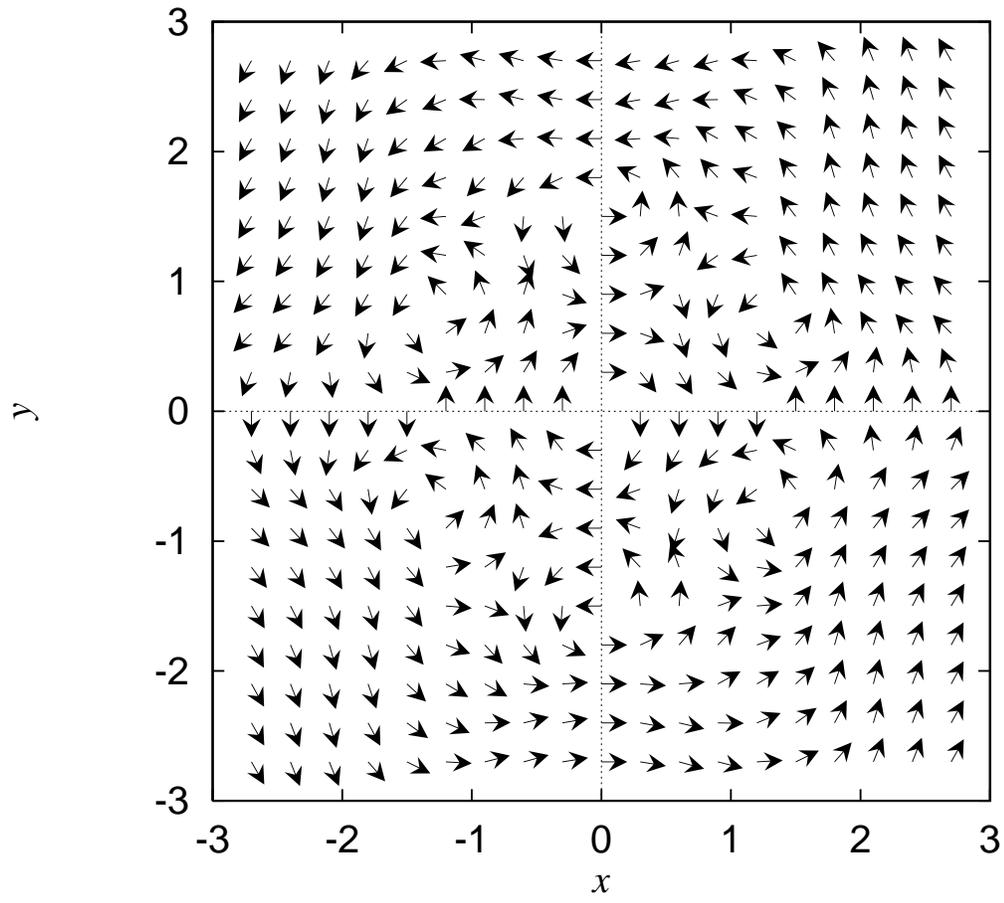
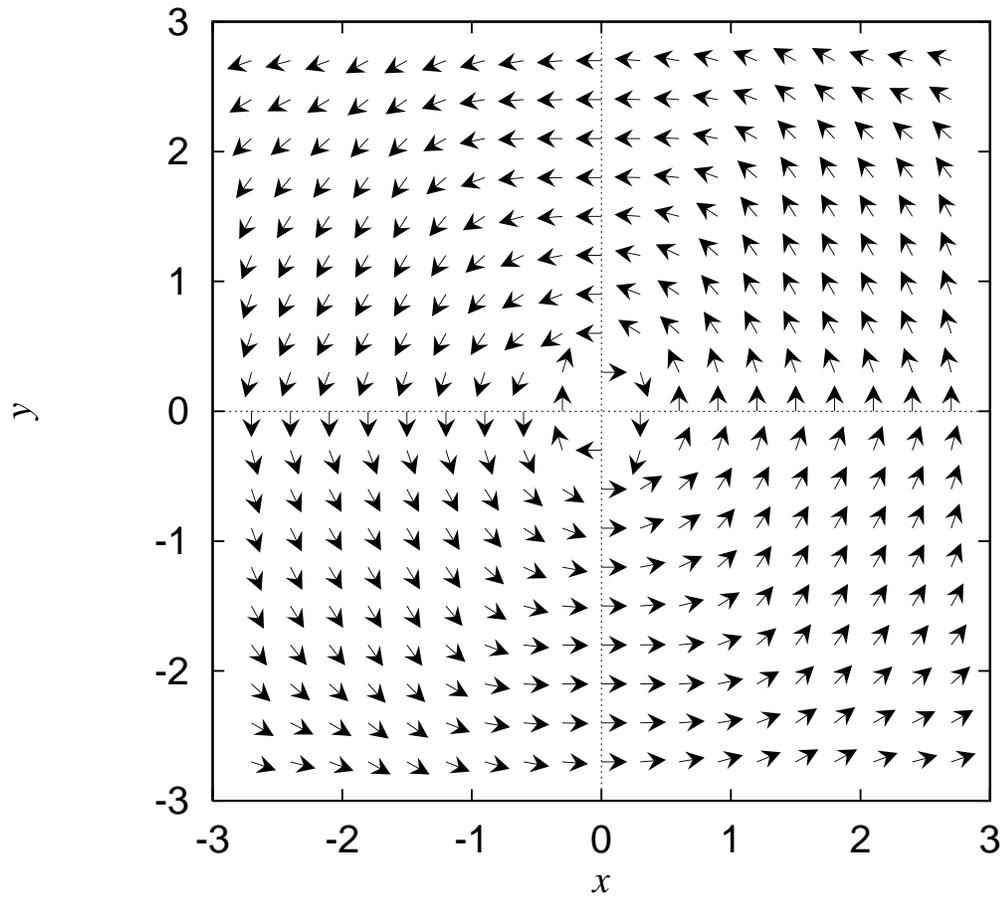
\begin{figure}
\caption{
The directional LDOS ${\bf I}(E,{\bf r})$
($\eta=0.03$)
for the energies $E=0.2$ (a), 1.2 (b), 1.4 (c), and 1.6 (d).
The arrows in the figures
represent only the directions
of ${\bf I}(E,{\bf r})$.
}
\label{fig:arrow}
\end{figure}


\begin{references}
\bibitem{scatt}
Strictly speaking, the states for $\theta$ in the vicinity of nodes are in the
scattering state,
and $N(E,{\bf r},\theta)$ is spread on the real space for these $\theta$.
\end{references}
\end{document}